\documentclass[a4paper,11pt]{article}
\usepackage[utf8]{inputenc}
\usepackage[margin=0.8in]{geometry}

\usepackage{amsmath}
\usepackage{amssymb}
\usepackage{bm}

\usepackage{graphicx}
\usepackage{pgfplots}

\usepackage{caption}
\usepackage{subfigure}

\usepackage{siunitx}

\usepackage{booktabs}

\usepackage[
colorlinks=true,
linkcolor=black,
anchorcolor=black,
citecolor=blue,
filecolor=magenta,
menucolor=red,
urlcolor=cyan,
plainpages=false,
pdfpagelabels,
hypertexnames=false,
linktocpage
]{hyperref}

\title{Numerical simulation of non-isothermal viscoelastic flows at high Weissenberg numbers using a finite volume method on general unstructured meshes}
\author{\small Stefanie Meburger$^{1,3}$, Matthias Niethammer$^{2}$\footnote{niethammer@mma.tu-darmstadt.de} , Dieter Bothe$^{2,3}$ and Michael Sch{\"a}fer$^{1,3}$\\ \\ 
\footnotesize
$^{1}$ Institute of Numerical Methods in Mechanical Engineering, TU Darmstadt\\
\footnotesize
$^{2}$ Institute of Mathematical Modeling and Analysis, TU Darmstadt\\
\footnotesize
$^{3}$ Graduate School of Computational Engineering, TU Darmstadt\\
\footnotesize
Dolivostr. 15, 64293 Darmstadt, Germany}

\begin{document}

\maketitle

\begin{abstract}

In this numerical study, an original approach to simulate non-isothermal viscoelastic fluid flows at high Weissenberg numbers is presented. Stable computations over a wide range of Weissenberg numbers are assured by using the root conformation approach in a finite volume framework on general unstructured meshes. The numerical stabilization framework is extended to consider thermo-rheological properties in Oldroyd-B type viscoelastic fluids.

The temperature dependence of the viscoelastic fluid is modeled with the time-temperature superposition principle. Both Arrhenius and WLF shift factors can be chosen, depending on the flow characteristics.
The internal energy balance takes into account both energy and entropy elasticity. Partitioning is achieved by a constant split factor.

An analytical solution of the balance equations in planar channel flow is derived to verify the results of the main field variables and to estimate the numerical error. 
The more complex entry flow of a polyisobutylene-based polymer solution in an axisymmetric 4:1 contraction is studied and compared to experimental data from the literature. We demonstrate the stability of the method in the experimentally relevant range of high Weissenberg numbers. The results at different imposed wall temperatures, as well as Weissenberg numbers, are found to be in good agreement with experimental data. 
Furthermore, the division between energy and entropy elasticity is investigated in detail with regard to the experimental setup.\\[5mm]
\textit{Keywords:} Non-isothermal; Viscoelastic; Entry flow; Finite Volume; Root conformation; Thermal effects

\end{abstract}

\section{Introduction}

The non-isothermal character of viscoelastic fluids is an important property 
when regarding their flow behavior.
In many industrial applications, such as polymer processing, viscoelastic flows are subject to thermal effects. Large temperature gradients occur in the fluid due to heating or cooling of the walls, while thermal conductivity and heat transfer are low~\cite{bird1987dynamics}. A considerable amount of mechanical energy is locally converted to thermal energy and the flow field is altered~\cite{bird1995}.
Numerical simulations can provide a deeper insight into these complex flow mechanisms and help to gain a better understanding and improvement of the process, e.g.~\cite{spanjaards2020}.

The temperature dependence of linear viscoelastic properties can be included in the constitutive equation by using the time-temperature superposition principle~\cite{bird1995}. This principle assumes that all model relaxation times vary with temperature in the same way, described by a shift factor~\cite{williams1955}. Two empirical descriptions of the shift factor are widely used: the William-Landel-Ferry (WLF) and Arrhenius approach. The use for a specific test-case depends on the modeled fluid and the temperature range~\cite{ferry1980}.

Special care needs to be taken for conversion mechanisms of internal energy. Showing both viscous and elastic behavior, the thermal energy is partly dissipated and partly stored in the fluid. Two ways of storing elastic energy have been found: entropy and internal energy elasticity~\cite{braun1990}. The exact conversion mechanism is complex, anisotropic and depends on the local flow behavior. For its description, at least an additional internal structural variable would be needed~\cite{huetter2009}. Braun~\cite{braun1991} established the idea of a constant weighting factor that describes the ratio of entropy to energy elasticity. This description facilitates the balance equation and has been taken up by Peters and Baaijens~\cite{peters1997} to develop an internal energy equation for multiple rate-type fluids. The concept has been adopted by subsequent numerical studies in the literature (cf.~\cite{wachs2000,wachs2002,habla2012}) and we will also rely on this approach.

Only limited rheological data on non-isothermal viscoelastic fluids are available in the literature that can be used for validation. Analysis of these fluids is difficult, models that describe ``real'' fluid behavior are complex and often imply many modes. Yet a comparison of simulation data to experimental data is indispensable to assure the validity of the used models. In the experimental study that we refer to, a highly elastic polyisobutylene-based polymer solution (PIB-Boger fluid) was investigated~\cite{yesilata2000}.
The Boger fluid is an artificial fluid developed to simplify experimental analysis and to close the gap between experimental observation and numerical prediction~\cite{boger1987}. Its viscosity is nearly constant over a wide range of flow rates such that the flow behavior can be described by a simple rate type model~\cite{boger1985}. For the simulation of the PIB-Boger fluid, we choose the Oldroyd-B model.

Flow at high elasticity, i.e. at high Weissenberg number, is of practical importance (cf.~\cite{shaw2012}), yet difficult to simulate numerically. Numerical solutions tend to become unstable at increased Weissenberg numbers, referred to as the High Weissenberg Number Problem (HWNP). To cope with the HWNP, various stabilization methods for viscoelastic solvers have been developed.
A common way to stabilize the computation is to introduce an additional diffusive term in the momentum balance equation, for instance with both sides diffusion~\cite{xue1995} or DEVSS \cite{guenette1995}. 
While stabilizing the calculation, the additional diffusive term tends to develop ``over-diffusion'' and does not seem to be suitable for transient flow~\cite{xue2004}.
A more sophisticated approach for stabilization is to solve a constitutive equation for an auxiliary variable instead of the polymeric stress tensor. This idea goes back to Fattal and Kupferman~\cite{fattal2005} who proposed a transport equation for the logarithm of the conformation tensor. Balci et al.~\cite{balci2011} developed a similar method with the square root of the conformation tensor as an auxiliary variable and we will use a related approach.

The objective of this work is threefold: (1) to present an extended stabilization method for simulating non-isothermal viscoelastic flows under experimentally realistic conditions; (2) to study complex entry flows at high Weissenberg numbers and predict thermo-rheological flow features, such as viscous dissipation; (3) to investigate the influence of the energy splitting factor in the limit of pure energy elasticity and pure entropy elasticity.
The new numerical framework is verified by comparison to analytical data and validated with experimental data from the literature.

The paper is organized as follows: in the next two sections, the thermo-rheological and the numerical model are described. 
In the consecutive section, an analytical solution for the field variables \textit{velocity}, \textit{first normal stress} and \textit{temperature} in planar channel flow is derived with constant viscosity and relaxation time. Analytical solutions are compared to simulation data in order to verify the functioning of the code. Mesh convergence and numerical errors are investigated.
Section \ref{comparison} describes the setup of a numerical test case that mimics the experiments performed by Yesilata et al.~\cite{yesilata2000} and discusses the results of the validation at different temperatures and Weissenberg numbers. Additionally, the influence of the splitting factor is investigated.
The last section summarizes the previous results.

\section{Mathematical model}
\label{sec:mathmodel}
The dynamics of the viscoelastic fluid are described by the incompressible continuity and momentum balance equation. The complex fluid behavior is modeled using the solvent-polymer stress splitting model SPSS proposed by Bird et al.~\cite{bird1987dynamics}. The stress tensor is split into a Newtonian solvent $\bm{\tau}_{\text{s}}$ and a polymeric part $\bm{\tau}_{\text{p}}$ according to
\begin{equation}
 \bm{\tau} = \bm{\tau}_{\text{s}} + \bm{\tau}_{\text{p}}.
\end{equation}
We choose the Oldroyd-B model for description of the polymeric stress tensor with the constitutive equation
\begin{equation}
 \bm{\tau}_{\text{p}} + \lambda \overset{\triangledown}{\bm{\tau}_{\text{p}}} = 2 \eta_{\text{p}} \mathbf{D} .
\label{OldroydBeq}
\end{equation}
Here $\overset{\triangledown}{\bm{\tau}_{\text{p}}}$ is the upper convected time derivative and $\mathbf{D} = \frac{1}{2} \left[ \bm{\nabla} \mathbf{u} + (\bm{\nabla} \mathbf{u})^{\sf T}\right] $ the deformation rate tensor.
For comparative computations, the exponential Phan-Thien-Tanner (PTT) model~\cite{phanthien1977} with the constitutive equation
\begin{equation}
 \text{exp}\left(\frac{ \lambda \epsilon}{\eta_{\text{p}}} \text{tr}(\bm{\tau}_{\text{p}})  \right) \bm{\tau}_{\text{p}} 
+ \lambda \overset{\triangledown}{\bm{\tau}_{\text{p}}} 
= 2 \eta_{\text{p}} \mathbf{D} 
\end{equation}
is employed. Here, $\epsilon$ is a material parameter related to the fluid behavior in extensional flow. The following modeling of temperature dependence and the energy equation are applicable to both rheological models.\\
The time-temperature superposition principle~\cite{ferry1980} is employed to describe the non-isothermal behavior of the fluid. Values of the model parameters solvent and polymeric viscosities $\eta_{\text{s}}$, $\eta_{\text{p}}$ and relaxation time $\lambda$ at a specific temperature are related to values at a reference temperature by the temperature-dependent shift factor $a_\text{T}(T)$. For the highly elastic polyisobutylene-based polymer solution used in the experiment we refer to, an Arrhenius approach is best suited to calculate $a_\text{T}(T)$. The reference temperature and the activation energy are given in \cite{yesilata2000} as
\begin{equation}
 \frac{\eta_{\text{s}}(T)}{\eta_{\text{s0}}} = \frac{\eta_{\text{p}}(T)}{\eta_{\text{p0}}} = \frac{\lambda(T)}{\lambda_\text{0}} = a_\text{T}(T).
\end{equation}
Here, $a_\text{T}(T)$ is the Arrhenius shift factor, $\eta_{\text{s0}}$, $\eta_{\text{p0}}$ and $\lambda_\text{0}$ are reference values of viscosities and relaxation time at reference temperature $T_\text{0}$. The shift factor is further dependent on the activation energy $\Delta H$ and the universal gas constant $R$ according to
\begin{equation}
  a_\text{T}(T) = exp \left[ \frac{\Delta H}{R} \left( \frac{1}{T} - \frac{1}{T_0} \right)\right].
	\label{arrheniusshift}
\end{equation}
The internal energy balance equation is the basis for the temperature transport equation. For viscoelastic fluids, the internal energy is a function of strain and temperature, leading to the heat equation (adapted from Peters and Baaijens \cite{peters1997})
\begin{equation}
  \frac{\partial ( \rho c_\text{p} T )}{\partial t} + \mathbf{u} \cdot \bm{\nabla} \left( \rho  c_\text{p} T \right) + \bm{\nabla} \cdot \mathbf{q} = Q .
	\label{energyeq}
 \end{equation}
Here $\rho$ denotes the density, $c_\text{p}$ the specific heat capacity, $\mathbf{q}$ the heat flux and $Q$ the energy source term. Fourier's law is employed to describe the heat conduction according to $\mathbf{q}=-k \bm{\nabla} T$ with the thermal conductivity $k$. The source term includes the thermal energy that results from conversion of mechanical energy and accounts for viscous dissipation and elastic storage. For an exact modeling of the ratio of dissipated to stored energy, at least one additional structural variable would be needed \cite{huetter2009}, yet this would go beyond the scope of the present study. Peters and Baaijens \cite{peters1997} proposed instead a pre-defined uniform splitting factor $\alpha$ and we will follow this approach. With these simplifications, the source term is found to be
\begin{equation}
  Q = \bm{\tau}_{\text{s}} : \mathbf{D} + \alpha \bm{\tau}_{\text{p}} : \mathbf{D} + (1 - \alpha) \frac{tr(\bm{\tau}_{\text{p}})}{2 \lambda(T)}.
	\label{tempequation}
 \end{equation}
The two limiting cases are $\alpha=0$, referred to as pure energy elasticity, where all converted energy is stored as elastic energy and can be released again and $\alpha=1$, referred to as pure entropy elasticity, where all energy is irreversibly dissipated.

\section{Numerical model}
The numerical model is implemented into a well-proven and robust FV framework for viscoelastic flows at high Weissenberg numbers, which has been used in previous works for isothermal single-phase~\cite{niethammer2018, niethammer2019} and two-phase~\cite{niethammer2019b, niethammer2019c} flows. The reader is referred to Niethammer et al.~\cite{niethammer2018, niethammer2019c} for a detailed description of the numerical discretization, the implementation of the root conformation approach and the velocity-stress coupling on co-located FV meshes. The FV framework for viscoelastic fluids is used on top of the open-source library OpenFOAM~\cite{weller1998}, which includes fully parallelized second-order FV schemes and iterative solvers for systems of linear equations.

In this work, we further extend the FV framework to solve the thermo-rheological model described in section \ref{sec:mathmodel}. The balance equations for momentum, stress and temperature are implemented into a segregated solution procedure. This section describes the implementation of the non-isothermal solver and summarizes the key aspects of the underlying FV framework for viscoelastic fluids.

\subsection{Numerical stabilization}
The numerical stabilization of differential constitutive stress equations, such as the Oldroyd-B equation (\ref{OldroydBeq}), is crucial in most CFD applications to avoid the High-Weissenberg number problem (HWNP)~\cite{joseph1985, keunings1986}. The HWNP refers to the breakdown of numerical computations at certain degrees of fluid elasticity, characterized by a critical problem-dependent Weissenberg number. A lack of convergence due to the HWNP is reported in the literature for all numerical methods used in computational rheology. Although the HWNP is not yet rigorously solved, effective stabilization methods are available. Fattal and Kupferman~\cite{fattal2004} showed that a logarithmic change of variables circumvents the high Weissenberg number instability.
Balci et al.~\cite{balci2011} proposed a square root conformation tensor representation that does not require any diagonalization of the conformation tensor. Detailed computational benchmark studies in an isothermal 4:1 contraction~\cite{niethammer2018} suggest that change-of-variable representations with small root functions show a better mesh-convergence compared to the logarithm conformation representation. Therefore, we choose the 4th root function, aiming to achieve a good compromise between stability and mesh-convergence. The root conformation tensor representation of the Oldroyd-B model can be written as
\begin{equation}
\label{RCROldroydB}
\partial_t \mathbf{R}
+ \left( \mathbf{u} \cdot \bm{\nabla} \right) \mathbf{R}
=
\frac{2}{k} \mathbf{B} \cdot \mathbf{R}
+
\boldsymbol{ \Omega } \cdot {\mathbf{R}} - {\mathbf{R}} \cdot \boldsymbol{ \Omega }
+
\frac{1}{k \lambda} \left(\mathbf{R}^{1-k} - \mathbf{R}\right),
\end{equation}
where $\mathbf{R}$ is the $k$-th root of the symmetric and positive definite conformation tensor $\mathbf{C}$.
The relation to the polymer stress is given by
\begin{equation}
\label{stressconformation}
\bm{\tau}_{\text{p}} = \frac{\eta_{\text{p}}}{\lambda} \left( \mathbf{C} - \mathbf{I} \right),
\end{equation}
where $\mathbf{I}$ is the unit tensor. The tensor variable $\mathbf{R}$ is computed from the diagonalization of the conformation tensor $\mathbf{C} = \mathbf{Q} \cdot \boldsymbol{ \Lambda } \cdot {\mathbf{Q}^{\top}}$ with the diagonal tensor $\boldsymbol{ \Lambda }$, containing the three real eigenvalues and the orthogonal tensor $\mathbf{Q}$, which includes the corresponding set of eigenvectors. For the inverse transformation $\mathbf{R}^k = \mathbf{C}$,
no diagonalization is used. Moreover, the convective derivative is decomposed into the first three terms on the r.h.s.~of (\ref{RCROldroydB}), containing the tensors $\mathbf{B}$ and $\boldsymbol{ \Omega }$. This local decomposition was first proposed in \cite{fattal2004}. The tensor $\mathbf{B}$ can be computed as $\mathbf{B} = \mathbf{Q} \cdot \tilde{\mathbf{B}} \cdot {\mathbf{Q}^{\top}}$, where the elements of the diagonal tensor $\tilde{\mathbf{B}}$ are given as a function of the tensor ${\bm{\nabla}\mathbf{u}^{\top}} = \mathbf{L} = \mathbf{Q} \cdot \tilde{\mathbf{L}} \cdot {\mathbf{Q}^{\top}}$ as $\tilde{b}_{ii} = \tilde{l}_{ii}$. The tensor $\boldsymbol{ \Omega }$ can be computed as $\boldsymbol{ \Omega } = \mathbf{Q} \cdot \tilde{\boldsymbol{ \Omega }} \cdot {\mathbf{Q}^{\top}}$, where the tensor $\tilde{\boldsymbol{ \Omega }}$ has zero diagonal entries $\tilde{\omega}_{ii} = 0$, while its off-diagonal elements are given by 
\begin{equation}
\label{diagonalizeCr82}
\tilde{\omega}_{{{ij}, \; {i \neq j}}} = \frac{\lambda_{ii} {\tilde{l}_{{ij}, \; {i \neq j}}} + \lambda_{jj} {\tilde{l}_{{ji}, \; {j \neq i}}}}{\lambda_{jj} - \lambda_{ii}}, \ i, j = 1, 2, 3.
\end{equation}
The generic numerical framework proposed in~\cite{niethammer2018} facilitates the construction and solution of certain stabilized representations of the form (\ref{RCROldroydB}). The generic procedure for assembling and solving the constitutive equations can be summarized in 4 steps:

\begin{enumerate}
	\item Construct the transport variable $\mathbf{R}$, using (\ref{stressconformation}) and a diagonalization of the conformation tensor. Here, the eigenvalues and eigenvectors are computed by a QL algorithm for symmetric matrices, based on Bowdler et al.~\cite{bowdler1971} and the corresponding routines in EISPACK.
	\item Decompose $\mathbf{L}$ into $\mathbf{B}$ and $\boldsymbol{ \Omega }$.
	\item Solve (\ref{RCROldroydB}), using second-order finite volume discretization schemes and an iterative method for the algebraic equation system.
	\item Transform $\mathbf{R}$ to $\boldsymbol{\tau}_{\text{p}}$, using the tensor product $\mathbf{R}^k = \mathbf{C}$ and (\ref{stressconformation}). No diagonalization is applied in the back transformation.
\end{enumerate}

\subsection{Discretization and velocity-stress coupling}
The finite volume method on general unstructured meshes is used for numerical discretization. The implementation is done on top of the widely used open-source package OpenFOAM~\cite{weller1998}, which provides a wide range of second-order finite volume schemes. A detailed description of the discretization practice is given in~\cite{niethammer2018} for the viscoelastic model and in~\cite{darwish2015} for the standard schemes in OpenFOAM. The time discretization is accomplished by using an implicit second-order Adams-Moulton scheme. High Resolution (HR) schemes in the Total Variation Diminishing (TVD) formulation~\cite{sweby} are employed for the discretization of convection terms. The van Leer flux limiter~\cite{vanLeer1974} is used in the HR schemes for all convection terms. For the constitutive tensor equation, a deferred correction (DC) procedure is used, where the coefficients of the higher-order interpolation are inserted as a source term in the algebraic equation. We choose the DC procedure because of its better stability, compared to the standard TVD implementation.

In a FVM with co-located variable arrangement, the velocity-stress coupling must be addressed similarly as proposed by Rhie and Chow~\cite{rhiechow1983} for the pressure-velocity coupling to prevent unphysical checkerboarding solutions in the flow fields. The velocity-stress coupling is considered by a correction term in the momentum equation as proposed by Niethammer et al.~\cite{niethammer2018}.
The correction removes the decoupling between the velocity and stress fields within our FVM on a general unstructured mesh. For the special case in this work, the correction $\mathbf{c}$ reduces to an anisotropic diffusion term which is added to the momentum equation, leading to
\begin{equation}
\mathbf{c}
=
\label{correction01}
\bm{\nabla} \cdot \left(\bm{\Gamma} \cdot \bm{\nabla} \mathbf{u} \right)
-
\overline{\bm{\nabla} \cdot \left(\bm{\Gamma} \cdot \bm{\nabla} \mathbf{u} \right)}.
\end{equation}
The key aspect of this approach lies in the different discretization of the two additional diffusion terms, such that the difference eliminates the cell-face interpolation errors caused by the discretization of the stress divergence. Because of the different discretization, one term is denoted by an overbar and $\mathbf{c}$ is not a zero addition. The diffusion tensor $\bm{\Gamma}$ can be computed from the matrix coefficients of the stress equation
\begin{equation}
\label{salgebraicform01}
a_P \bm{\tau}_P + \sum_N a_N \bm{\tau}_N = \mathbf{S}_P,
\end{equation}
where $\mathbf{S}_P$ represents the discretized source or sink terms and the coefficients $a_P$ and $a_N$ for a cell-centered point $P$ and its neighbors $N$ read
\begin{align}
\label{coeffaps01a}
a_P = \frac{3 V_P}{2 \Delta t} + a_P^{adv}, \quad
a_N = a_N^{adv}.
\end{align}
The first term in $a_P$ results from the discretization of the temporal term with the time step $\Delta t$ and the cell volume $V_P$. The advection parts $a_P^{adv}$ and $a_N^{adv}$ in the coefficients depend on the high resolution scheme and the van Leer flux limiter. Using the coefficient $a_P$, the diffusion tensor can be written as
\begin{equation}
\bm{\Gamma}
=
\frac{1}{a_P} \left(
\bm{\tau}_P + \frac{\eta_p}{\lambda} \mathbf{I}
\right).
\end{equation}

\subsection{Solution algorithm}
The problem of non-isothermal viscoelastic flow is solved iteratively by a segregated approach. The procedure used to solve the coupled set of equations can be summarized in the following steps:
\begin{enumerate}
	\item \textit{Initialization.} For given initial fields of $p$, $\eta_{\text{s}}(T)$, $\eta_{\text{p}}(T)$, $\lambda(T)$, $T$, $\mathbf{u}$, $\boldsymbol{\tau}_{\text{p}}$ and the generic tensor transport variable $\mathbf{R}$, compute a cell-centroid velocity estimate from the discretized momentum equation.
	\item \textit{SIMPLE algorithm.} Solve the pressure equation implicitly and, subsequently, correct the cell-face fluxes. Update the velocity, using the new pressure gradient.
	\item \textit{Constitutive equation.} Assemble the constitutive equation (\ref{stressconformation}), using the new velocity. Compute the new constitutive transport variable $\mathbf{R}$ by solving the constitutive equation implicitly and, subsequently, update the stress $\boldsymbol{\tau}_{\text{p}}$.
	\item \textit{Temperature equation.} Assemble the temperature equation (\ref{tempequation}), using the new velocity and the new stress.
	\item \textit{Update the fluid properties.} Compute the Arrhenius shift factor (\ref{arrheniusshift}), using the new temperature field and update the fields $\eta_{\text{s}}(T)$, $\eta_{\text{p}}(T)$, $\lambda(T)$.
	\item \textit{Optionally repeat (only for transient solutions).} Repeat steps 1 to 5 within each time step to increase the accuracy of the transient solution.
\end{enumerate}
Within this procedure, the discretized systems of linear equations are solved by iterative methods.
A conjugate gradient method with algebraic multigrid preconditioning is used for the pressure. A bi-conjugate gradient stabilized method with incomplete lower-upper preconditioning is used for stress and temperature.

\section{Non-isothermal channel flow}
In order to verify the new numerical framework, the analytical solution for an Oldroyd-B fluid with constant viscosities and relaxation time is calculated and compared to numerical results. Three fluid field variables are compared: axial velocity, first normal stress component and temperature.
\subsubsection*{Analytical solution}
The axial velocity profile for an Oldroyd-B fluid in a plane channel reads
\begin{equation}
 u(y)=\frac{3}{2} \overline{u} \left( 1 - \frac{y^2}{H^2} \right) .
\end{equation}
The profile of the first normal component of the stress tensor is found to be
\begin{equation}
 \bm{\tau}_{\text{p,xx}}(y) = \frac{18 \lambda \eta_{\text{p}} \overline{u}^2}{H^4} {y}^{2} .
\end{equation}
In the following section, the analytical profile of temperature for an Oldroyd-B fluid with constant properties is deduced.
The starting point is the energy equation for non-isothermal, viscoelastic fluids (\ref{energyeq}). 
A steady state is considered, so that temporal derivatives and derivatives in axial direction do not play any role. Due to the conservation of mass, the velocity perpendicular to the axial direction is zero. In fully-developed pure shear flow, all internal energy is dissipated and the splitting parameter $\alpha$ can be assumed equal to one \cite{wapperom1998thermo}. With the stated assumptions, the heat equation reduces to
\begin{equation}
 k \frac{\partial^2 T}{\partial y^2} + \left( \bm{\tau}_{\text{s,xy}} + \bm{\tau}_{\text{p,xy}} \right) \frac{\partial u_{\text{x}}}{\partial y} = 0 .
\label{energyeqreduced}
\end{equation}
Equation (\ref{energyeqreduced}) is integrated, taking into account the boundary conditions $T(y=H)=T_w$ and $\frac{\partial T}{\partial x}(y=0)=0$. The temperature profile for steady, planar channel flow is found to be
\begin{equation}
 T ( y) = - \frac{3}{4} \frac{\text{Br*}}{H^4} y^4 + \frac{3}{4} \text{Br*} + T_{\text{w}} 
\end{equation}
with a variation of the Brinkman number $\text{Br*} = \frac{\overline{u}^2 \eta_{\text{0}}}{k} $.

\subsubsection*{Numerical setup}
The fluid properties for the results in this section are given by the density $\rho=\SI{921} {\kg\per\cubic\m}$, the viscosity $\eta_{\text{0}}=\SI{1e4} {Pa~s}$ with a ratio of solvent to polymer viscosity of $1/19$, the specific heat $c_{\text{p}}=\SI{1500} {\J\per\kg\per\K}$ and the thermal conductivity $k=\SI{0.13}{\W\per\m\per\K}$.

The calculations are performed on four different meshes, generated by gradually increasing the number of grid cells perpendicular to the flow direction $\text{N}_\text{dy}$ from 10 to 40. At the inlet, Dirichlet boundary conditions are assumed for temperature and velocity $T_{\text{in}}=\SI{462}{\K}$, $\overline{u_{\text{x,in}}}=\SI{0.01}{\m\per\s}$. A zero normal derivative is imposed for stress tensor and pressure. At the wall, no-slip boundary conditions are assumed for the velocity and Dirichlet boundary conditions are employed for the temperature with $T_{\text{w}}=\SI{462}{\K}$. At the outlet, all variables are imposed to have zero normal derivative except for a fixed pressure value.

The relative error $\delta_x$, measuring the deviation of the calculated solution from the analytical solution, is defined as follows
\begin{equation}
\delta_x = \frac{\tilde{x}-x}{x-x_0} ,
\end{equation}
where $x$ is the exact solution, $x_0$ the initial value and $\tilde{x}$ the approximated value of the variable $x$.
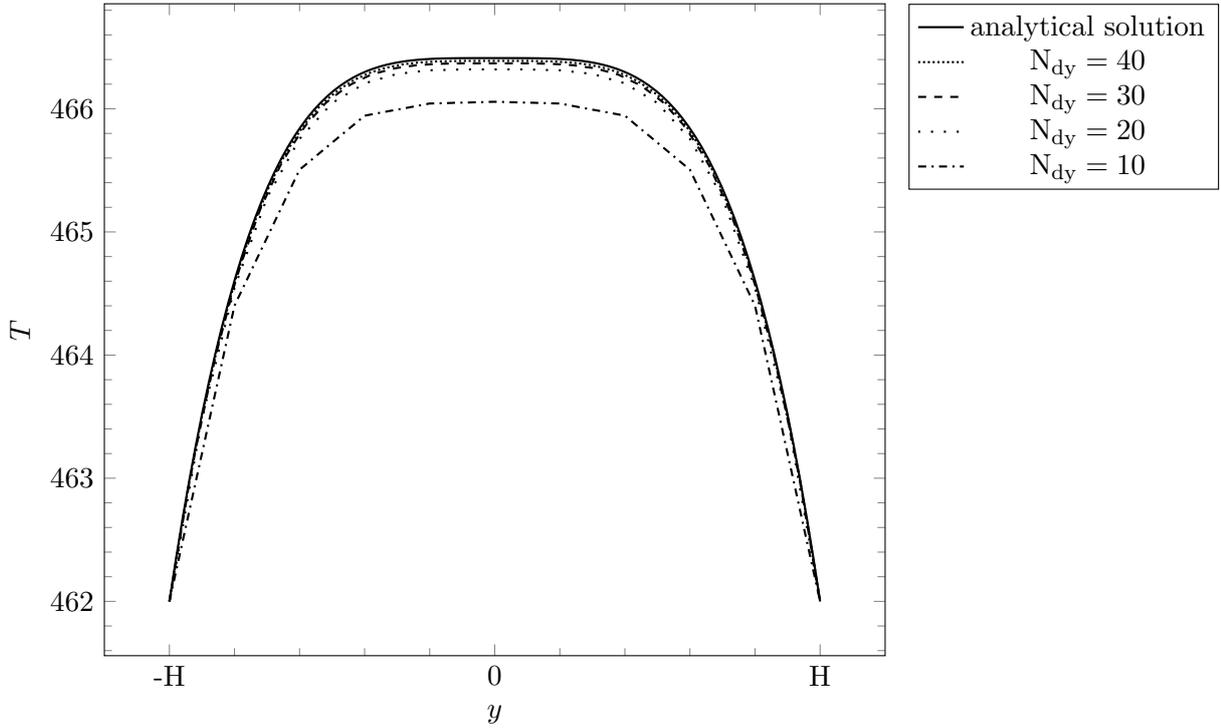
\begin{figure}[ht!]
 \begin{tikzpicture}
 \begin{axis}[
	width=0.7\textwidth,
  xlabel={$y$},       
	ylabel={$T$},
	xlabel near ticks,
	ylabel near ticks,
	minor y tick num=4,
	scaled ticks=false,
	minor x tick num=4,
	legend pos=outer north east,
	xtick={-0.001,0,0.001},
	xticklabels={-H,0,H},
	]
	\addplot [thick] table {Data/462_output_t.csv};
	\addlegendentry{analytical solution}
	\addplot [thick,densely dotted] table {Data/an02gr05_t_x04_t60.csv};
	\addlegendentry{$\text{N}_\text{dy}=40$}
	\addplot [thick,dashed] table {Data/an02gr04_t_x04_t60.csv};
	\addlegendentry{$\text{N}_\text{dy}=30$}
	\addplot [thick,loosely dotted] table {Data/an9bm_t_x03_t60.csv};
	\addlegendentry{$\text{N}_\text{dy}=20$}
	\addplot [thick,dashdotted] table {Data/an7_t_x03_t60.csv};
	\addlegendentry{$\text{N}_\text{dy}=10$}	
 \end{axis} 
 \end{tikzpicture}
 \captionof{figure}{Temperature profile as a function of the channel height}
 \label{figure:tprofile}
\end{figure}
Figure \ref{figure:tprofile} shows the analytical temperature profile as a function of the channel height in comparison to numerical solutions on all considered meshes. Deviations are visible for the bulk temperature and reduce with increasing mesh refinement.
\begin{figure}%
\centering
\subfigure[][]{%
  \label{fig:relerr-a}%
	\begin{tikzpicture}
   \begin{axis}[
	width=0.45\textwidth,
  xlabel={$\text{N}_\text{dy}$},       
	ylabel={$\delta_{u_{\text{x}}}$},
	xlabel near ticks,
	ylabel near ticks,
	ymin=0, ymax=11,
	minor x tick num=1,
	minor y tick num=3,
	]
	\addplot [only marks,mark=+] table {Data/relerr_u_an7_gr05.csv};
  \end{axis} 
 \end{tikzpicture}}%
\hspace{8pt}%
\subfigure[][]{%
 \label{fig:relerr-b}%
  \begin{tikzpicture}
  \begin{axis}[
	width=0.45\textwidth,
  xlabel={$\text{N}_\text{dy}$},       
	ylabel={$\delta_{\bm{\tau}_{\text{xx}}}$},
	xlabel near ticks,
	ylabel near ticks,
	ymin=0, ymax=11,
	minor x tick num=1,
	minor y tick num=3,
	]
	\addplot [only marks,mark=+] table {Data/relerr_tau_an7_gr05.csv};
  \end{axis} 
  \end{tikzpicture}}\\
	\subfigure[][]{%
	 \label{fig:relerr-c}
    \begin{tikzpicture}
    \begin{axis}[
	width=0.45\textwidth,
  xlabel={$\text{N}_\text{dy}$},       
	ylabel={$\delta_{T}$},
	xlabel near ticks,
	ylabel near ticks,
	ymin=0, ymax=11,
	minor x tick num=1,
	minor y tick num=3,
	legend pos=outer north east,
	]
	\addplot [only marks,mark=+] table {Data/relerr_t_an7_gr05.csv};
   \end{axis} 
   \end{tikzpicture}}%
	\caption{Relative error $\delta$ in $\%$ as a function of the grid resolution in y-direction $\text{N}_\text{dy}$ for
	\subref{fig:relerr-a} velocity $u_{\text{x}}$
	\subref{fig:relerr-b} stress tensor component $\bm{\tau}_{\text{xx}}$ and
	\subref{fig:relerr-c} temperature $T$}
	\label{figure:relerr}%
\end{figure}
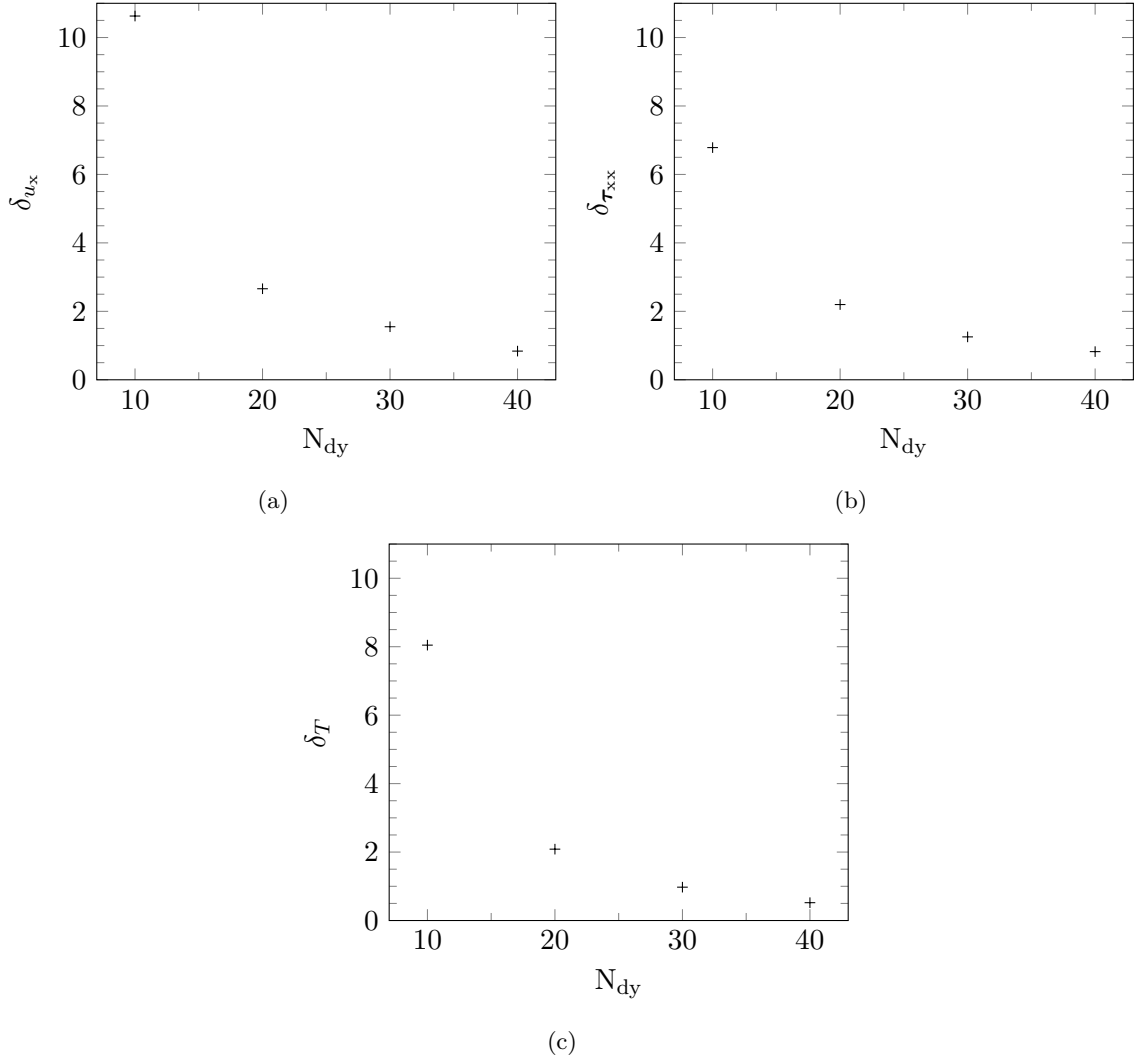
In Figure \ref{figure:relerr} the relative error for axial velocity, first normal stress component and temperature is plotted as a function of the number of grid cells in y-direction $\text{N}_\text{dy}$. For all variables, the numerical error reduces quadratically, thus a second-order mesh convergence is achieved.

\section{Comparison to experimental data}
\label{comparison}
\subsection{Setup of the test case}
\begin{figure}
  \begin{tikzpicture}
  \tikzstyle{every node}=[font=\footnotesize]
   \begin{axis}[
   hide axis,
   enlargelimits=false,
   x=0.95\columnwidth,
   y=0.5\columnwidth,
   xmin=0.0,xmax=1.1,ymin=-0.15,ymax=0.6,
   ]
    \draw[dash pattern=on 0.4pt off 3pt on 4pt off 3pt] (axis cs:0,0) -- (axis cs:1,0);
    \draw[dotted] (axis cs:0.15,0) -- (axis cs:0.15,0.4);
    \draw[thick, line cap=rect] (axis cs:0.15,0.4) -- (axis cs:0.2,0.4);
    \draw[line width=1pt, line cap=rect] (axis cs:0.2,0.4) -- (axis cs:0.7,0.4);
    \draw[line width=1pt, line cap=rect] (axis cs:0.7,0.4) -- (axis cs:0.7,0.1);
    \draw[line width=1pt, line cap=rect] (axis cs:0.7,0.1) -- (axis cs:0.9,0.1);
    \draw[dotted] (axis cs:0.9,0.1) -- (axis cs:0.9,0);
    \draw (axis cs:0.07,0.4) -- (axis cs:0.13,0.4);
    \draw[->] (axis cs:0.1,0) -- (axis cs:0.1,0.4);
    \draw (axis cs:0.15,0.42) -- (axis cs:0.15,0.48);
    \draw (axis cs:0.2,0.42) -- (axis cs:0.2,0.48);
    \draw[<->] (axis cs:0.15,0.45) -- (axis cs:0.2,0.45);
    \draw (axis cs:0.7,0.42) -- (axis cs:0.7,0.48);
    \draw[<->] (axis cs:0.2,0.45) -- (axis cs:0.7,0.45);
    \draw (axis cs:0.9,0.42) -- (axis cs:0.9,0.48);
    \draw[<->] (axis cs:0.7,0.45) -- (axis cs:0.9,0.45);
    \draw (axis cs:0.92,0.1) -- (axis cs:0.98,0.1);
    \draw[->] (axis cs:0.95,0) -- (axis cs:0.95,0.1);
    \node[left] at (axis cs:0.1,0.2) {$4R_{\text{2}}$};
    \node[right] at (axis cs:0.95,0.05) {$R_{\text{2}}$};
    \node[above] at (axis cs:0.175,0.45) {$4R_{\text{2}}$};
    \node[above] at (axis cs:0.45,0.45) {$92R_{\text{2}}$};
    \node[above] at (axis cs:0.8,0.45) {$24R_{\text{2}}$};
    \draw (axis cs:0.175,0.395) -- (axis cs: 0.18,0.3);
    \node[below] at (axis cs:0.19,0.3) {$T_{\text{in}}$};
    \draw (axis cs:0.15,0.15) -- (axis cs:0.18,0.2);
    \node[below]  at (axis cs:0.45,0.39) {$T_{\text{w}}$};
    \node[left]  at (axis cs:0.7,0.25) {$T_{\text{w}}$};
    \node[above]  at (axis cs:0.8,0.1) {$T_{\text{w}}$};
    \draw[thick,->] (axis cs:0.3,0.1) -- (axis cs:0.35,0.1);
    \node[above] at (axis cs: 0.325,0.1) {$u$};    
    \draw[thick,->] (axis cs:0.7,0) -- (axis cs:0.7375,0);
    \draw[thick,->] (axis cs:0.7,0) -- (axis cs:0.7,0.075);
    \node[below] at (axis cs:0.71,0) {\scriptsize $x$};
    \node[left] at (axis cs:0.7,0.035) {\scriptsize $r$};
   \end{axis}
  \end{tikzpicture}
  \captionof{figure}{2D sketch of the axisymmetric test case geometry}
  \label{figure:sketch_domain}
 \end{figure}
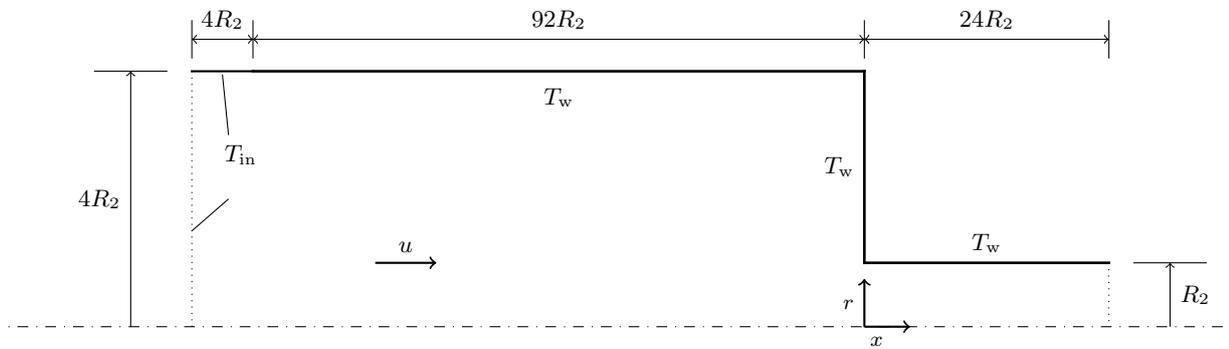
 
Figure \ref{figure:sketch_domain} shows the geometry of the axisymmetric domain. Its shape is modeled as close as possible to the experimental flow domain investigated by Yesilata et al.~\cite{yesilata2000}. The contraction ratio is 4:1, radii and length of the inlet and outlet ducts as well as the length of the heated/cooled wall are equal to the experimental setup. The radius of the outlet duct is $R_{\text{2}}=\SI{6.35}{mm}$.
At the inlet, uniform values are given for velocity and temperature; the stress tensor is imposed to have zero normal derivative.
Inlet velocities are predefined by the respective Weissenberg numbers. Inlet temperature is $\SI{296.5}{\K}$, the reference room temperature given by Yesilata et al.~\cite{yesilata2000}.

For the velocity field, no-slip boundary conditions are imposed. Dirichlet boundary conditions are set for temperature at the walls. The wall temperature of the first wall segment is equal to the inlet temperature, the consecutive wall segments are heated or cooled as indicated in the respective results.

At the outlet, the pressure is fixed, while all other field variables follow a zero normal derivative. In Table \ref{table:fluidprop}, the fluid properties of the highly elastic polyisobutylene-based polymer solution are given as reported in \cite{yesilata2000}. The values of viscosities and relaxation time correspond to a limit of zero shear-rate at reference temperature of $\SI{296.5}{\K}$.

While the numerical setup is chosen as close as possible to the experiment, some differences are present and should be explained. Firstly, instead of simulating the whole circular pipe, only an axisymmetric pipe segment is calculated. In order to investigate the possible error due to the imposed symmetry in the circumferential direction, some of the simulations are also performed in a fully three-dimensional half-cylinder.
The other difference is the inlet of the pipe. In the experiment, the fluid enters the observation domain from a smaller pipe of unknown radius. In the numerical setup, a uniform velocity is imposed at the inlet. As a consequence, velocity, stress and temperature profiles vary in the vicinity of the inlet. However, the inlet duct is long enough to allow the profiles to fully develop and the differences at the inlet are not assumed to affect the investigated flow behavior in the vicinity of the contraction.
 
\begin{table}[t]
\centering
 \begin{tabular}{lc|clc}
 \toprule
  $\rho$  & \SI{880}{\kg\per\cubic\m} & $c_{\text{p}}$ & \SI{1970}{\J\per\kg\per\K}\\[1.1mm]
  $\eta_{\text{s0}}$  & \SI{31}{\Pa\s} &  $\eta_{\text{p0}}$  & \SI{17}{\Pa\s}\\[1.1mm]
  $\lambda_{\text{0}}$  & \SI{2.0}{\s} &   $T_{\text{0}}$ & \SI{296.5}{\K}\\[1.1mm]
  $k$  & \SI{0.13}{\W\per\m\per\K} &   $\frac{\Delta H}{R}$  &\SI{6414}{\K} \\
  \bottomrule
 \end{tabular}
\captionof{table}{Fluid properties of the highly elastic polyisobutylene-based polymer solution}
\label{table:fluidprop}
\end{table}
\begin{figure}
\centering
\includegraphics[width=0.99\textwidth,trim=1cm 12cm 1cm 12cm,clip]{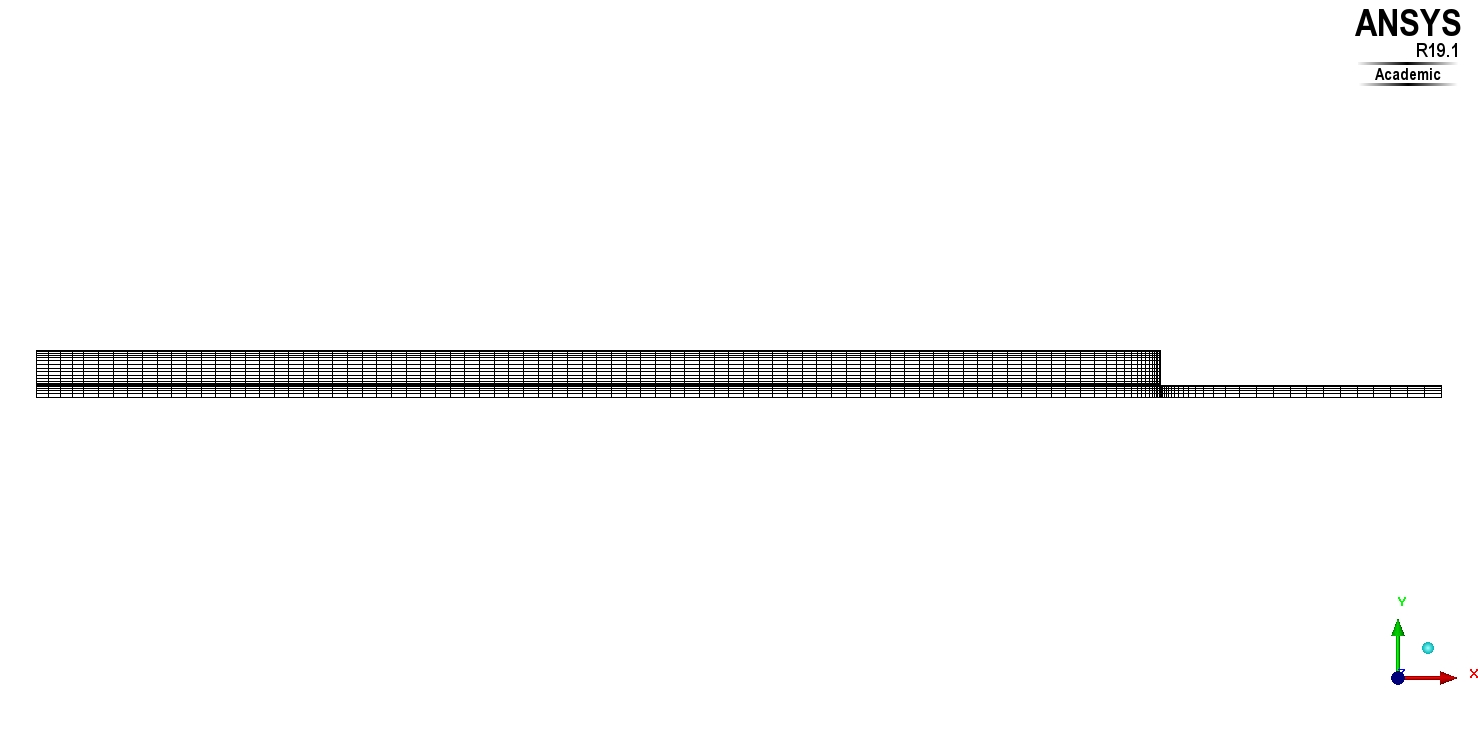}
\caption{Visualization of the coarsest mesh 1}
\label{fig:coarsemesh}
\end{figure}
The grid sensitivity is investigated by using three stepwise refined meshes, whose number of control volumes and ratio of the smallest cell to outlet duct radius are shown in Table \ref{table:meshparam}. In Figure \ref{fig:coarsemesh}, the coarsest mesh is depicted.

\begin{table}
\centering
 \begin{tabular}{ccc}
 \toprule
 Grid & Control volumes & $\Delta x_{min}/R_{\text{2}}$ \\
 \midrule
 mesh 1 & 1800 & 0.096 \\
 mesh 2 & 5560 & 0.048 \\
 mesh 3 & 21500 & 0.024\\
 half-cylinder & 46540 & 0.096\\
 \bottomrule
\end{tabular}
\captionof{table}{Mesh parameters}
\label{table:meshparam}
\end{table}

Three dimensionless numbers play an important role to describe the complex fluid dynamics. The Reynolds Number $Re = \frac{\overline{u_{\text{x,2}}} R_{\text{2}} \rho}{\eta_{\text{0}}}$ measures the ratio of inertial to viscous forces. The value is calculated in the outlet duct, with mean axial velocity $\overline{u_{\text{x,2}}}$ and total viscosity $\eta_{\text{0}}=\eta_{\text{s}}+\eta_{\text{p}}$.
The Weissenberg number $Wi$ describes the ratio of elastic to viscous forces in viscoelastic materials, defined as $Wi = \frac{\lambda \overline{u_{\text{x,2}}}}{R_{\text{2}}}$. The Deborah Number $De$ is defined as the ratio of characteristic time of the fluid to the time scale of the process. In steady flow, as considered in this study, it is equal to the Weissenberg number and both are used equivalently here.

The contraction is the origin of the coordinate system as illustrated in Figure \ref{figure:sketch_domain}. The axial coordinate is non-dimensionalized with the outlet duct radius, i.e. $\zeta = x / R_{\text{2}}$. Negative values of $\zeta$ refer to the inlet duct, positive values to the outlet duct. The presented results have been evaluated (if not stated otherwise) slightly upstream of the contraction, at the axial position $\zeta = - 0.3$.

The temperature field data is presented in dimensionless form of $\theta = \frac{T_{\text{w}} - T}{T_{\text{w}} - T_{\text{in}}}$ with inlet temperature $T_{\text{in}}=\SI{296.5}{K}$ and respective wall temperature of the heated/cooled wall $T_{\text{w}}$.

\subsection{Results and discussion}
Calculations are performed at three different wall temperatures: a cooled wall of $T_{\text{w}} = \SI{285}{\K}$ and heated walls of $T_{\text{w}} = \SI{305}{\K}$ and $T_{\text{w}} = \SI{327}{\K}$. The results of the computations are presented in this section and compared to experimental data, reproduced from Yesilata et al.~\cite{yesilata2000}. The measurements were taken across the entire cylinder, however all values are shown in the positive radial direction here.

\subsubsection*{Results without stabilization}
For comparison, a simulation is performed using a solution procedure without any kind of numerical stabilization. The calculation is tested with wall temperature $T_{\text{w}} = \SI{327}{\K}$ at Weissenberg number $Wi=12.3$ on the finest mesh 3. We observe an abrupt crash of the simulation with a floating-point exception error after a simulation time of about $\SI{7}{\s}$. A possible reason for the numerical breakdown could be the HWNP.

All further results presented in this chapter are calculated with the root conformation approach, for which we did not encounter any instability issues.

\subsubsection*{Effect of the imposed circumferential symmetry}
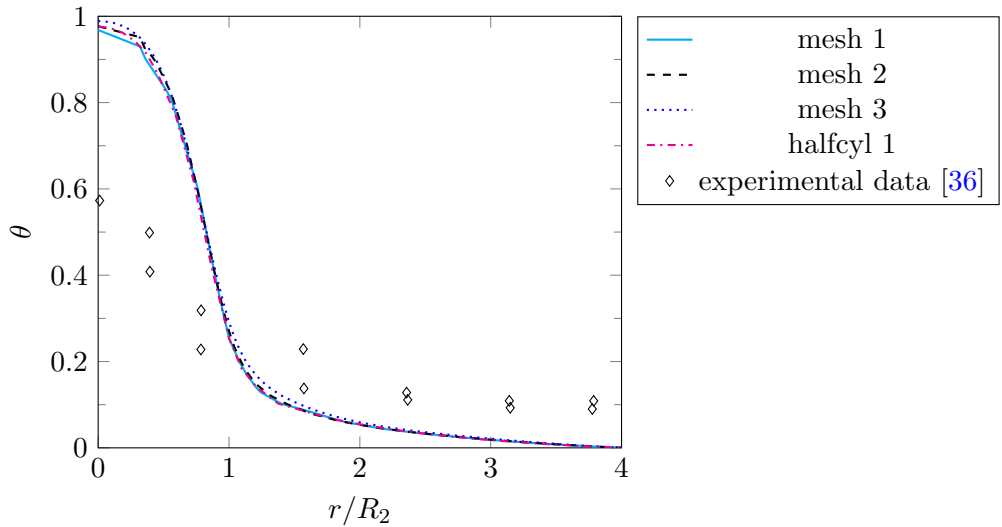
\begin{figure}
  \begin{tikzpicture}
  \begin{axis}[
	legend pos=outer north east,
	width=0.5\textwidth,
  xlabel={$r/R_2$},       
	ylabel={$\theta$},
	xlabel near ticks,
	ylabel near ticks,	
	ymin=0,
	ymax=1,
	xmin=0,
	xmax=4,
	minor y tick num=1,
	]	
 	\addplot [cyan,thick] table {Data/theta_exp05_dt285_de504_al05_t1200.csv};
 	\addlegendentry{mesh 1}
	\addplot [black, thick,dashed] table {Data/theta_exp04_dt285_de504_al05_t1200.csv};
	\addlegendentry{mesh 2}
	\addplot [blue, thick,dotted] table {Data/theta_exp03_dt285_de504_al05_t1200.csv};
	\addlegendentry{mesh 3}
	\addplot [magenta, thick,dashdotted] table {Data/theta_hc05_dt285_de504_al05_t1200.csv};
	\addlegendentry{halfcyl 1}
	\addplot [mark=diamond, only marks, black] table {Data/exp_yesi285_de504_near_01.csv};
	\addlegendentry{experimental data \cite{yesilata2000}}
	\addplot [mark=diamond, only marks, black] table {Data/exp_yesi285_de504_near_02.csv};
  \end{axis}
 \end{tikzpicture}
	\caption{Dimensionless temperature vs. radial position at $Wi = 5$, $\zeta=-0.3$ for $T_{\text{w}} = \SI{285}{\K}$}
	\label{figure:tw285hc}%
\end{figure}
The axisymmetry of the numerical setup enforces symmetry in the circumferential direction. As the numerical flow profile is symmetric in the radial and circumferential direction, this is not expected to affect the solution. In order to investigate if the assumption is valid, a fully three-dimensional simulation in a half-cylinder is performed for comparison. Details on the numerical grid can be found in Table \ref{table:meshparam}, the refinement of the wall boundary layers is mostly equivalent to mesh 1.

The dash-dotted line in Figure \ref{figure:tw285hc} shows the temperature profile of the three-dimensional simulation. The solid, dashed and dotted line show results of mesh 1, 2 and 3. 
Despite the higher number of degrees of freedom, the profile coincides with the two-dimensional numerical results. 
Circumferential flow is possible, yet minimal, probably due to the lack of disturbances, and the bulk temperature is not reduced. 

\subsubsection*{Near-wall behavior}
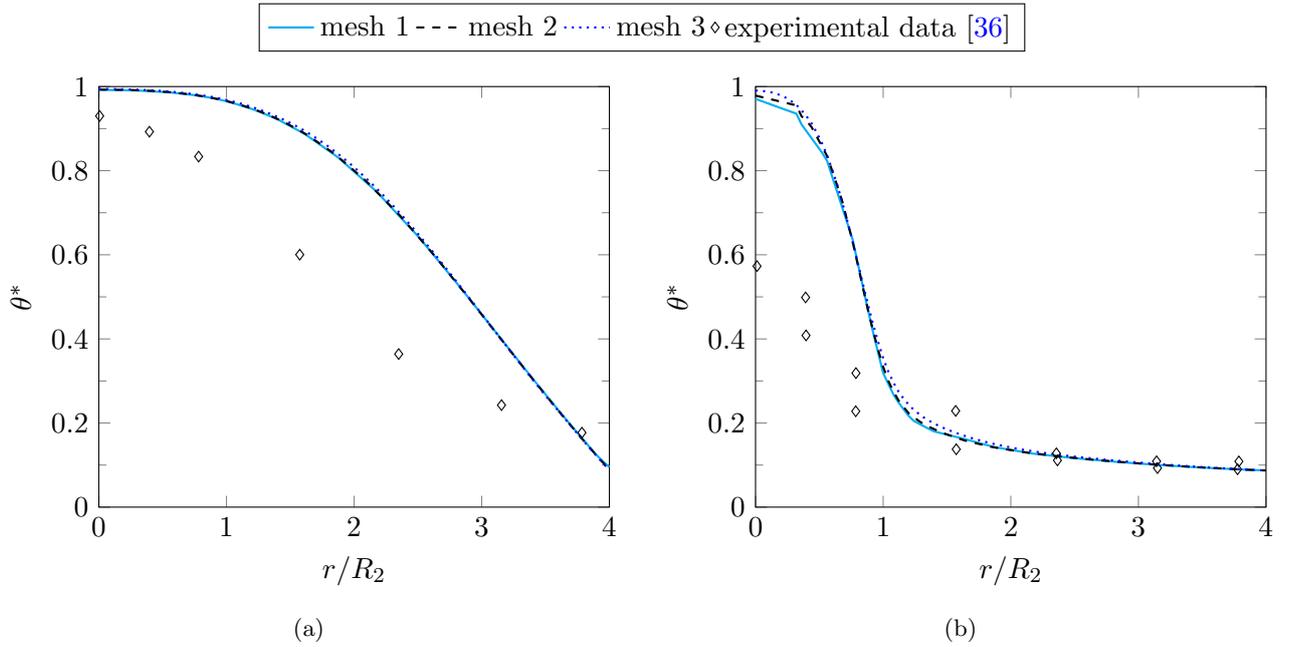
\begin{figure}%
\centering
\ref{zweifuenfacht}\\
\subfigure[][]{%
  \label{fig:tw285-a}%
  \begin{tikzpicture}
  \begin{axis}[
	legend style={legend columns=-1},
	legend to name=zweifuenfacht,
	width=0.49\textwidth,
  xlabel={$r/R_2$},       
	ylabel={$\theta^{*}$},
	xlabel near ticks,
	ylabel near ticks,	
	ymin=0,
	ymax=1,
	xmin=0,
	xmax=4,
	minor y tick num=1,
	]
	\addplot [cyan, thick] table {Data/theta_exp05_dt285_de504_al05_tw286_t1200_far.csv};
 	\addlegendentry{mesh 1}
	\addplot [black, thick,dashed] table {Data/theta_exp04_dt285_de504_al05_tw286_t1200_far.csv};
 	\addlegendentry{mesh 2}
	\addplot [blue, thick,dotted] table {Data/theta_exp03_dt285_de504_al05_tw286_t1200_far.csv};
 	\addlegendentry{mesh 3}
	\addplot [mark=diamond, only marks, black] table {Data/exp_yesi285_de504_far01.csv};
	\addlegendentry{experimental data \cite{yesilata2000}}
	\addplot [mark=diamond, only marks, black] table {Data/exp_yesi285_de504_far02.csv};
  \end{axis}
 \end{tikzpicture}}%
\hspace{8pt}%
\subfigure[][]{%
	 \label{fig:tw285-b}
  \begin{tikzpicture}
  \begin{axis}[
	width=0.49\textwidth,
  xlabel={$r/R_2$},       
	ylabel={$\theta^{*}$},
	xlabel near ticks,
	ylabel near ticks,	
	ymin=0,
	ymax=1,
	xmin=0,
	xmax=4,
	minor y tick num=1,
	]
 	\addplot [cyan, thick] table {Data/theta_exp05_dt285_de504_al05_tw286_t1200.csv};
 	\addplot [black, thick,dashed] table {Data/theta_exp04_dt285_de504_al05_tw286_t1200.csv};
	\addplot [blue, thick,dotted] table {Data/theta_exp03_dt285_de504_al05_tw286_t1200.csv};
	\addplot [mark=diamond, only marks, black] table {Data/exp_yesi285_de504_near_01.csv};
	\addplot [mark=diamond, only marks, black] table {Data/exp_yesi285_de504_near_02.csv};
  \end{axis}
 \end{tikzpicture}}%
	\caption{Dimensionless temperature vs. radial position for an adjusted setup with $T_{\text{w}} = \SI{286}{\K}$, $Wi = 5$ at the axial locations 
	\subref{fig:tw285-a} $\zeta=-8$ and
	\subref{fig:tw285-b} $\zeta=-0.3$}
	\label{figure:tw286}%
\end{figure}
 
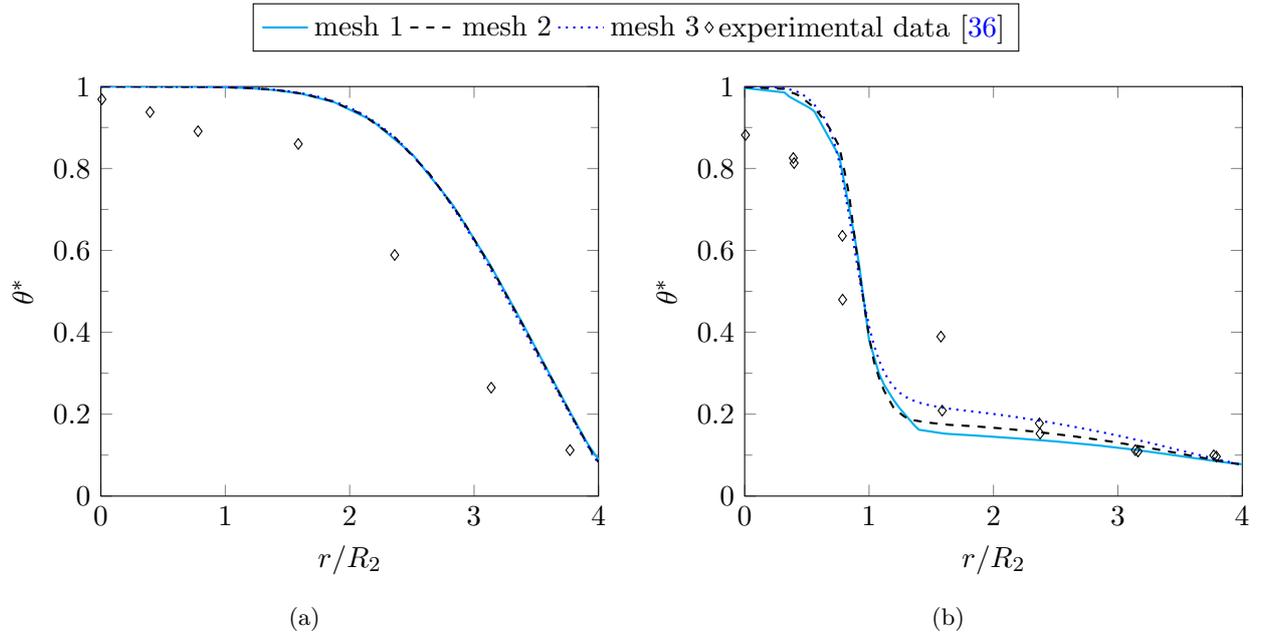
\begin{figure}[t]
\centering
\ref{dreinullfuenf}\\
\subfigure[][]{%
  \label{figure:305de113-a}%
  \begin{tikzpicture}
  \begin{axis}[
	legend style={legend columns=-1},
	legend to name=dreinullfuenf,
	width=0.48\textwidth,
  xlabel={$r/R_2$},       
	ylabel={$\theta^{*}$},
	xlabel near ticks,
	ylabel near ticks,	
	ymin=0,
	ymax=1,
	xmin=0,
	xmax=4,
	minor y tick num=1,
	]
 	\addplot [cyan, thick] table {Data/theta_exp05_dt305_de113_al05_tw30435_t1200_far.csv};
 	\addlegendentry{mesh 1}
 	\addplot [black, thick,dashed] table {Data/theta_exp04_dt305_de113_al05_tw30435_t800_far.csv};
 	\addlegendentry{mesh 2}
	\addplot [blue, thick,dotted] table {Data/theta_exp03_dt305_de113_al05_tw30435_t800_far.csv};
	\addlegendentry{mesh 3}
	\addplot [mark=diamond, only marks, black] table {Data/exp_yesi305_de113_far01.csv};
	\addlegendentry{experimental data \cite{yesilata2000}}
	\addplot [mark=diamond, only marks, black] table {Data/exp_yesi305_de113_far02.csv};
  \end{axis}
 \end{tikzpicture}}%
\hspace{8pt}%
\subfigure[][]{%
	 \label{figure:305de113-b}
	\begin{tikzpicture}
  \begin{axis}[
	width=0.48\textwidth,
  xlabel={$r/R_2$},       
	ylabel={$\theta^{*}$},
	xlabel near ticks,
	ylabel near ticks,	
	ymin=0,
	ymax=1,
	xmin=0,
	xmax=4,
	minor y tick num=1,
	]
 	\addplot [cyan, thick] table {Data/theta_exp05_dt305_de113_al05_tw30435_t1200.csv};
 	\addplot [black, thick,dashed] table {Data/theta_exp04_dt305_de113_al05_tw30435_t800.csv};
	\addplot [blue, thick,dotted] table {Data/theta_exp03_dt305_de113_al05_tw30435_t800.csv};
	\addplot [mark=diamond, only marks, black] table {Data/exp_yesi305_de113_near_01.csv};
	\addplot [mark=diamond, only marks, black] table {Data/exp_yesi305_de113_near_02.csv};
  \end{axis}
 \end{tikzpicture}
  }%
	\caption{Dimensionless temperature vs. radial position for an adjusted wall temperature of $T_{\text{w}} = \SI{304.35}{\K}$, $Wi = 11.3$  at
	\subref{figure:305de113-a} $\zeta=-8$ and
	\subref{figure:305de113-b} $\zeta=-0.3$}
	\label{figure:305de113}%
\end{figure}

Figure \ref{figure:tw285hc} shows the dimensionless temperature $\theta$ against the radial position $r/R_2$ for a cooled wall with temperature $T_{\text{w}} = \SI{285}{\K}$ at Weissenberg number $Wi = 5$. The solid, dashed and dotted curves show the temperature profiles resulting from calculations on the successively refined grids $1$, $2$ and $3$, respectively. 
While the qualitative shape of the temperature profile is captured well by the calculations, deviations can be observed especially at the wall and at the center-line.

At the wall ($r/R_2=4$), we have $\theta=0$ which corresponds to $T = T_\text{w}$. However, the value $\theta \approx 0.1$ is reported in~\cite{yesilata2000}, which corresponds to an increased wall temperature $T_\text{w} = \SI{286}{\K}$. This suggests that the measured temperature at the considered axial position differs slightly from the nominal wall temperature $T_{\text{w,nominal}}=\SI{285}{\K}$. For this reason, the simulations are adjusted to better agree with the experimentally reported wall temperature. Calculations with an adjusted setup, where the wall temperature is set to $T_{\text{w}} = \SI{286}{\K}$, are presented in the right picture of Figure \ref{figure:tw286}. The non-dimensionalization is still performed with the nominal wall temperature, according to $\theta^{*} = \frac{T_{\text{w,nominal}} - T}{T_{\text{w,nominal}} - T_{\text{in}}}$. Thus, $\theta^{*}$ is not zero at the wall but equals the wall temperature reported in the experiments.
Note that this adjustment does not alter the qualitative shape of the flow profiles but improves the consistency with the corresponding experimental temperature profile in the vicinity of the wall.
With the adjusted wall temperature, the temperature profiles near the wall and in the outer half of the cylinder are very close to the experiments.

In Figure \ref{figure:305de113}\subref{figure:305de113-b}, the dimensionless temperature $\theta^{*}$ calculated on the three meshes for a heated wall at Weissenberg number $Wi = 11.3$ is shown in the vicinity of the contraction at $\zeta=-0.3$. While the nominal wall temperature is $T_{\text{w,nominal}} = \SI{305}{\K}$, the simulations were performed with an adjusted setup at $T_{\text{w}} = \SI{304.35}{\K}$ as described above.
Also for the heated wall, the temperature profile in the vicinity of the contraction is found to be in good agreement with the experimental values in the outer half of the cylinder.

\subsubsection*{Bulk temperature}
Deviations between experimental data and simulation data are pronounced at the center-line ($r/R_2 = 0$). At this location, computed $\theta$ values are significantly larger than in the experimental data, and grid refinement tends to increase the deviation. In Figure \ref{figure:tw286}, the dimensionless temperature profile is shown at the two probe locations \subref{fig:tw285-a} $\zeta=-8$ and \subref{fig:tw285-b} $\zeta=-0.3$. Comparing the temperature values in the center at $r/R_2 = 0$, we observe only a slight decrease in temperature from \subref{fig:tw285-a} to \subref{fig:tw285-b} in the simulation. The experimental data are reported to decrease significantly at this location. Assuming the same value of nominal inlet temperature in the experiments and the simulation $T_{\text{in}}=\SI{296.5}{\K}$, the temperature drop in flow direction is significantly under-predicted in the simulation for the cooled test case. In the case of heated walls, shown in Figures \ref{figure:305de113} \subref{figure:305de113-a} and \subref{figure:305de113-b}, the increase of the bulk temperature is slightly smaller than in the experiments. Therefore, for both the heated and the cooled wall, the temperature change caused by the imposed wall temperatures is underestimated in the simulation.
As the thermal conductivity $k = \SI{0.13}{\W\per\m\per\K}$ is very low, the temperature increase or decrease is mainly due to heat production by viscous dissipation.
A prerequisite for viscous dissipation is the presence of a velocity gradient. In regions with pronounced velocity gradients, that is at the wall and in the re-circulation zone in front of the contraction, the temperature change due to viscous dissipation is large, and the temperature profile is captured well.
In the center of the cylinder, the velocity gradients are small, as is viscous dissipation. 
To explain the deviations, we need to recall the differences between numerical and experimental setup. The numerical calculations show ``ideal'' flow conditions where the symmetry of the velocity profiles in radial and circumferential direction is guaranteed. 
This symmetry is most unlikely in any natural flow, where small disturbances lead to enhanced secondary flow in the radial and the circumferential direction. 
The asymmetry of the flow profiles becomes evident through the two different values that were measured in the positive and negative radial direction. In perfect symmetry, both values would coincide.
In the cited experimental setup, it is also possible that additional secondary flow was created by the intrusion of the temperature probes. 
As a result, the viscous dissipation is expected to be larger in the experiments, causing a greater change of the bulk temperature. The calculated flow profile shows ``ideal'' flow conditions and could be regarded as a lower bound for converted energy with minimal viscous dissipation.

While the trend is clear, the magnitude of the stated deviations shall be calculated in absolute temperature values.
In the case of cooled walls, the highest difference in reported experimental values is at a radial position of $r/R_2 \approx 1.6$. The difference in dimensionless temperature $\Delta \theta$ is about $0.1$ which is equivalent to $\Delta T \approx \SI{1}{\K}$. The deviation of the bulk temperature is notably higher, about $\Delta \theta \approx 0.4$ equivalent to $\Delta T \approx \SI{4.8}{\K}$.
For the heated walls at $T_{\text{w}} = \SI{305}{\K}$, $\Delta \theta$ is about $0.2$. This corresponds to $\Delta T$ of about $\SI{1.5}{\K}$. The difference in temperature in the middle of the cylinder, where the deviation between numerical and experimental data is most pronounced, is $\Delta T \approx \SI{1}{\K}$ or $\Delta \theta \approx 0.1$ respectively.
To summarize, in the case of heated walls, the deviation between experimental and numerical data in the bulk temperature is of the same order of magnitude as the reported difference in the measured data at the same radial position. For cooled walls, the deviation in bulk temperature exceeds this difference.

\subsubsection*{Effect of the Weissenberg number}
\begin{figure}%
\centering
\ref{dreihundertsiebenundzwanzigfar}\\
\subfigure[][]{%
  \label{fig:tw327far-a}%
	\begin{tikzpicture}
  \begin{axis}[
	legend style={legend columns=-1},
	legend to name=dreihundertsiebenundzwanzigfar,
	width=0.48\textwidth,
  xlabel={$r/R_2$},       
	ylabel={$\theta$},
	xlabel near ticks,
	ylabel near ticks,	
	ymin=0,
	ymax=1,
	xmin=0,
	xmax=4,
	minor y tick num=1,
	]
	\addplot [magenta, thick,dashed] table {Data/theta_exp03_dt327_de40_al05_ptt_t1500_far.csv};
	\addlegendentry{mesh 3 PTT}
	\addplot [blue, thick,dotted] table {Data/theta_exp03_dt327_de40_al05_t1500_far.csv};
	\addlegendentry{mesh 3 Oldroyd-B}
	\addplot [mark=diamond, only marks, black] table {Data/exp_yesi327_de4_far01.csv};
	\addlegendentry{experimental data \cite{yesilata2000}}
	\addplot [mark=diamond, only marks, black] table {Data/exp_yesi327_de4_far02.csv};
  \end{axis}
 \end{tikzpicture}}%
\hspace{8pt}%
\subfigure[][]{%
 \label{fig:tw327far-b}%
  \begin{tikzpicture}
  \begin{axis}[
	width=0.48\textwidth,
  xlabel={$r/R_2$},       
	ylabel={$\theta$},
	xlabel near ticks,
	ylabel near ticks,	
	ymin=0,
	ymax=1,
	xmin=0,
	xmax=4,
	minor y tick num=1,
	]
	\addplot [magenta,thick,dashed] table {Data/theta_exp03_dt327_de64_al05_ptt_t1000_far.csv};
	\addplot [blue, thick,dotted] table {Data/theta_exp03_dt327_de64_al05_t1000_far.csv};
	\addplot [mark=diamond, only marks, black] table {Data/exp_yesi327_de64_far.csv};
  \end{axis}
 \end{tikzpicture}}\\
	\subfigure[][]{%
	 \label{fig:tw327far-c}
    \begin{tikzpicture}
  \begin{axis}[
	width=0.48\textwidth,
  xlabel={$r/R_2$},       
	ylabel={$\theta$},
	xlabel near ticks,
	ylabel near ticks,	
	ymin=0,
	ymax=1,
	xmin=0,
	xmax=4,
	minor y tick num=1,
	]
	\addplot [magenta, thick,dashed] table {Data/theta_exp03_dt327_de82_al05_ptt_t1000_far.csv};
	\addplot [blue, thick,dotted] table {Data/theta_exp03_dt327_de82_al05_t1000_far.csv};
	\addplot [mark=diamond, only marks, black] table {Data/exp_yesi327_de82_far01.csv};
	\addplot [mark=diamond, only marks, black] table {Data/exp_yesi327_de82_far02.csv};
  \end{axis}
 \end{tikzpicture}}%
\hspace{8pt}%
\subfigure[][]{%
	 \label{fig:tw327far-d}
\begin{tikzpicture}
  \begin{axis}[
	width=0.48\textwidth,
  xlabel={$r/R_2$},       
	ylabel={$\theta$},
	xlabel near ticks,
	ylabel near ticks,	
	ymin=0,
	ymax=1,
	xmin=0,
	xmax=4,
	minor y tick num=1,
	]
	\addplot [magenta,thick,dashed] table {Data/theta_exp03_dt327_de123_al05_ptt_t700_far.csv};
	\addplot [blue, thick,dotted] table {Data/theta_exp03_dt327_de123_al05_t700_far.csv};
	\addplot [mark=diamond, only marks, black] table {Data/exp_yesi327_de123_far01.csv};
	\addplot [mark=diamond, only marks, black] table {Data/exp_yesi327_de123_far02.csv};
  \end{axis}
 \end{tikzpicture}}%
	\caption{Dimensionless temperature vs. radial position at $T_w=\SI{327}{K}$, $\zeta=-8.0$ for
	\subref{fig:tw327far-a} $Wi=4.0$
	\subref{fig:tw327far-b} $Wi=6.4$
	\subref{fig:tw327far-c} $Wi=8.2$ and
	\subref{fig:tw327far-d} $Wi=12.3$}
	\label{figure:tw327de_far}%
\end{figure}
\begin{figure}%
\centering
\ref{dreihundertsiebenundzwanzignear}\\
\subfigure[][]{%
  \label{fig:tw327-a}%
	\begin{tikzpicture}
  \begin{axis}[
	legend style={legend columns=-1},
	legend to name=dreihundertsiebenundzwanzignear,
	width=0.48\textwidth,
  xlabel={$r/R_2$},       
	ylabel={$\theta$},
	xlabel near ticks,
	ylabel near ticks,	
	ymin=0,
	ymax=1,
	xmin=0,
	xmax=4,
	minor y tick num=1,
	]
	\addplot [magenta,thick,dashed] table {Data/theta_exp03_dt327_de40_al05_ptt_t1500.csv};
	\addlegendentry{mesh 3 PTT}
	\addplot [blue,thick,dotted] table {Data/theta_exp03_dt327_de40_al05_t1500.csv};
	\addlegendentry{mesh 3 Oldroyd-B}
	\addplot [mark=diamond, only marks, black] table {Data/exp_yesi327_de4_near01.csv};
	\addlegendentry{experimental data \cite{yesilata2000}}
	\addplot [mark=diamond, only marks, black] table {Data/exp_yesi327_de4_near02.csv};
  \end{axis}
 \end{tikzpicture}}%
\hspace{8pt}%
\subfigure[][]{%
 \label{fig:tw327-b}%
  \begin{tikzpicture}
  \begin{axis}[
	width=0.48\textwidth,
  xlabel={$r/R_2$},       
	ylabel={$\theta$},
	xlabel near ticks,
	ylabel near ticks,	
	ymin=0,
	ymax=1,
	xmin=0,
	xmax=4,
	minor y tick num=1,
	]
	\addplot [magenta,thick,dashed] table {Data/theta_exp03_dt327_de64_al05_ptt_t1000.csv};
	\addplot [blue,thick,dotted] table {Data/theta_exp03_dt327_de64_al05_t1000.csv};
	\addplot [mark=diamond, only marks, black] table {Data/exp_yesi327_de64_near01.csv};
	\addplot [mark=diamond, only marks, black] table {Data/exp_yesi327_de64_near02.csv};
  \end{axis}
 \end{tikzpicture}}\\
	\subfigure[][]{%
	 \label{fig:tw327-c}
    \begin{tikzpicture}
  \begin{axis}[
	width=0.48\textwidth,
  xlabel={$r/R_2$},       
	ylabel={$\theta$},
	xlabel near ticks,
	ylabel near ticks,	
	ymin=0,
	ymax=1,
	xmin=0,
	xmax=4,
	minor y tick num=1,
	]
	\addplot [magenta,thick,dashed] table {Data/theta_exp03_dt327_de82_al05_ptt_t1000.csv};
	\addplot [blue,thick,dotted] table {Data/theta_exp03_dt327_de82_al05_t1000.csv};
	\addplot [mark=diamond, only marks, black] table {Data/exp_yesi327_de82_near01.csv};
	\addplot [mark=diamond, only marks, black] table {Data/exp_yesi327_de82_near02.csv};
  \end{axis}
 \end{tikzpicture}}%
\hspace{8pt}%
\subfigure[][]{%
	 \label{fig:tw327-d}
\begin{tikzpicture}
  \begin{axis}[
	width=0.48\textwidth,
  xlabel={$r/R_2$},       
	ylabel={$\theta$},
	xlabel near ticks,
	ylabel near ticks,	
	ymin=0,
	ymax=1,
	xmin=0,
	xmax=4,
	minor y tick num=1,
	]	
	\addplot [magenta,thick,dashed] table {Data/theta_exp03_dt327_de123_al05_ptt_t700.csv};
	\addplot [blue,thick,dotted] table {Data/theta_exp03_dt327_de123_al05_t700.csv};
	\addplot [mark=diamond, only marks, black] table {Data/exp_yesi327_de123_near01.csv};
	\addplot [mark=diamond, only marks, black] table {Data/exp_yesi327_de123_near02.csv};
  \end{axis}
 \end{tikzpicture}}%
	\caption{Dimensionless temperature vs. radial position at $T_w=\SI{327}{\K}$, $\zeta=-0.3$ for
	\subref{fig:tw327-a} $Wi=4.0$
	\subref{fig:tw327-b} $Wi=6.4$
	\subref{fig:tw327-c} $Wi=8.2$ and
	\subref{fig:tw327-d} $Wi=12.3$}
	\label{figure:tw327de}%
\end{figure}
\begin{figure}%
\subfigure[][]{%
  \label{fig:tw327b-a}%
	\includegraphics[width=0.49\textwidth,trim=1cm 36cm 1cm 52.5cm,clip]{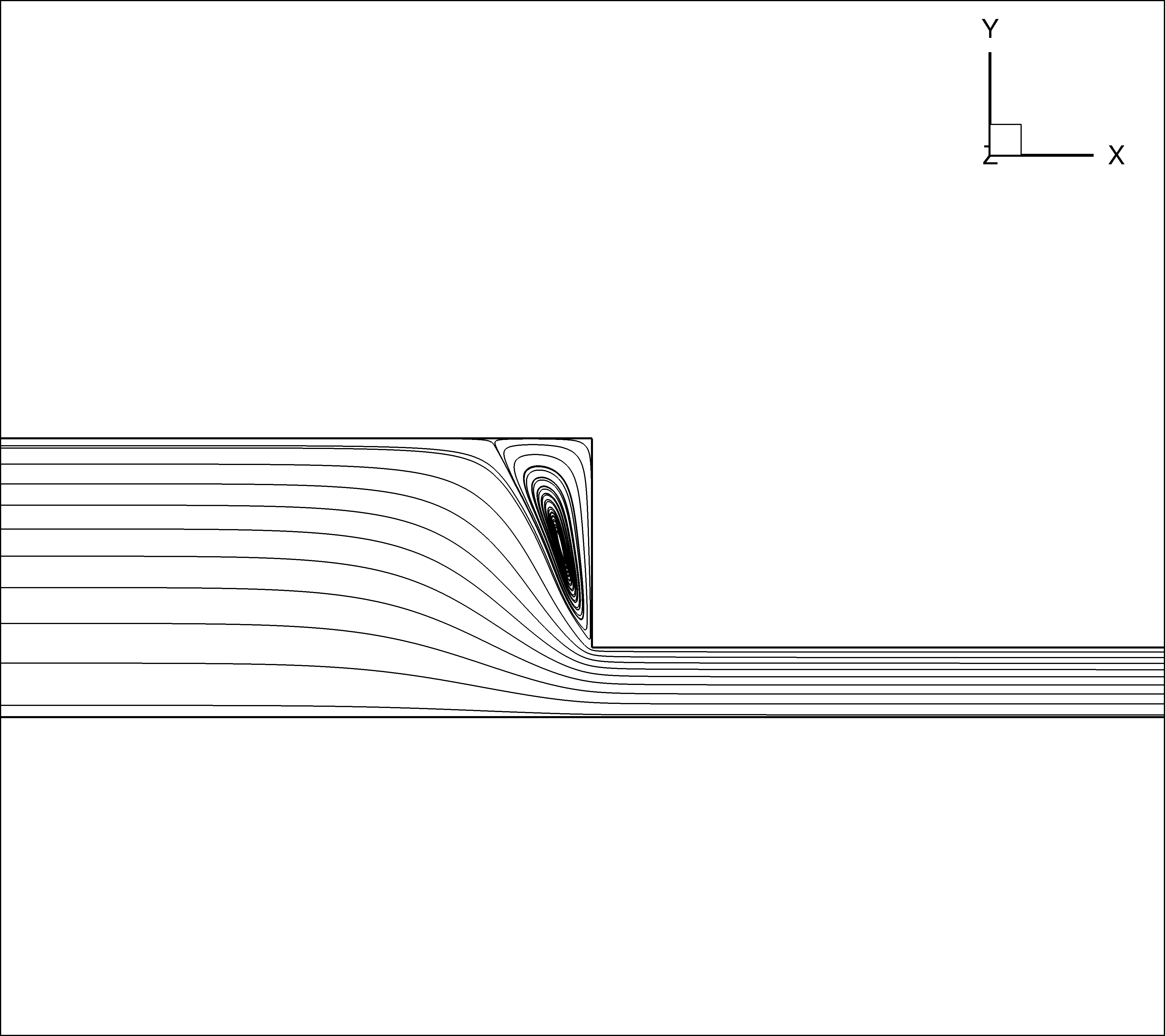}
	}%
\hspace{8pt}%
\subfigure[][]{%
 \label{fig:tw327b-b}%
\includegraphics[width=0.49\textwidth,trim=1cm 39cm 1cm 50cm,clip]{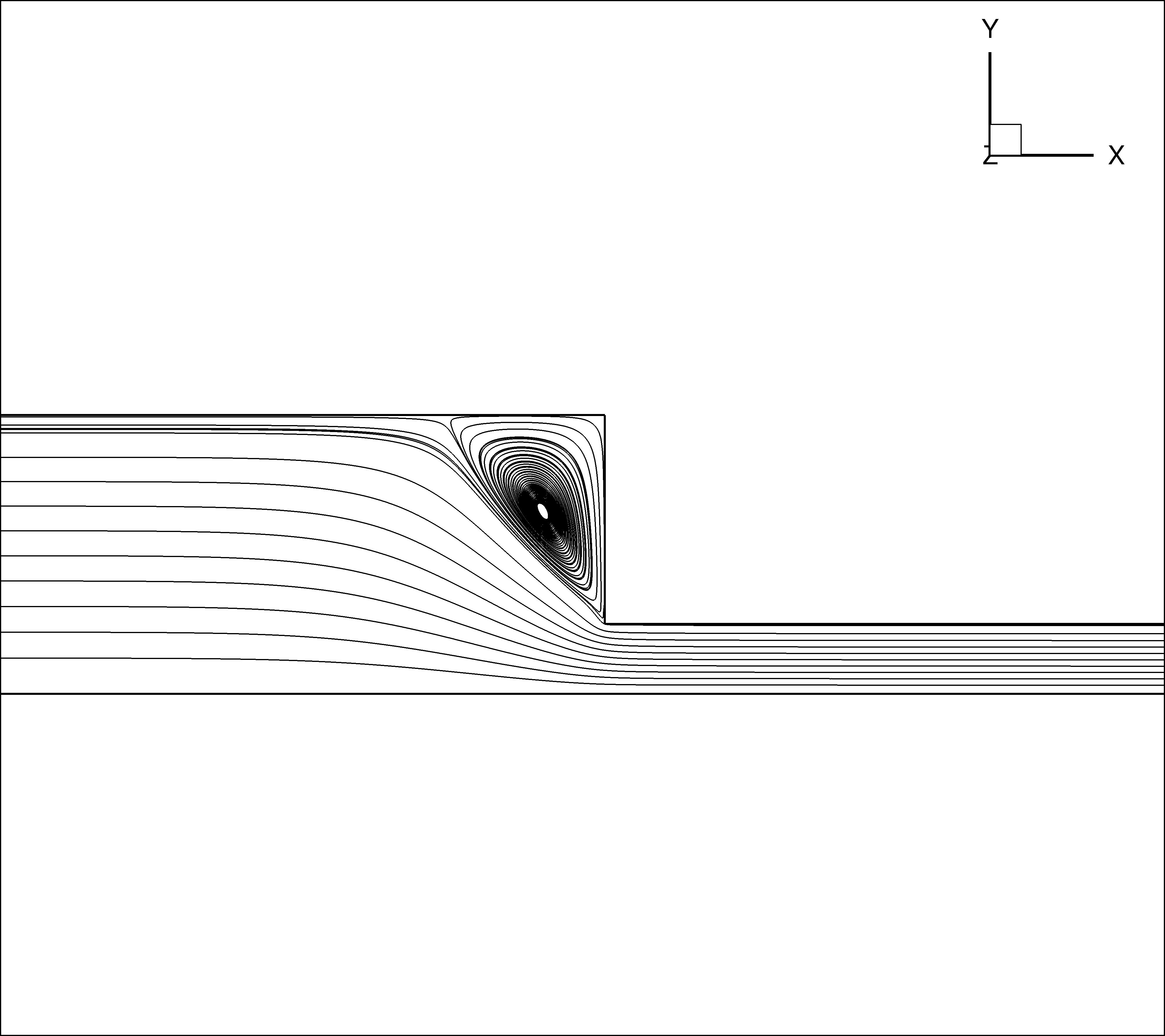}
  }\\
\subfigure[][]{%
  \label{fig:tw327b-c}%
	\includegraphics[width=0.49\textwidth,trim=10mm 30cm 10mm 50cm,clip]{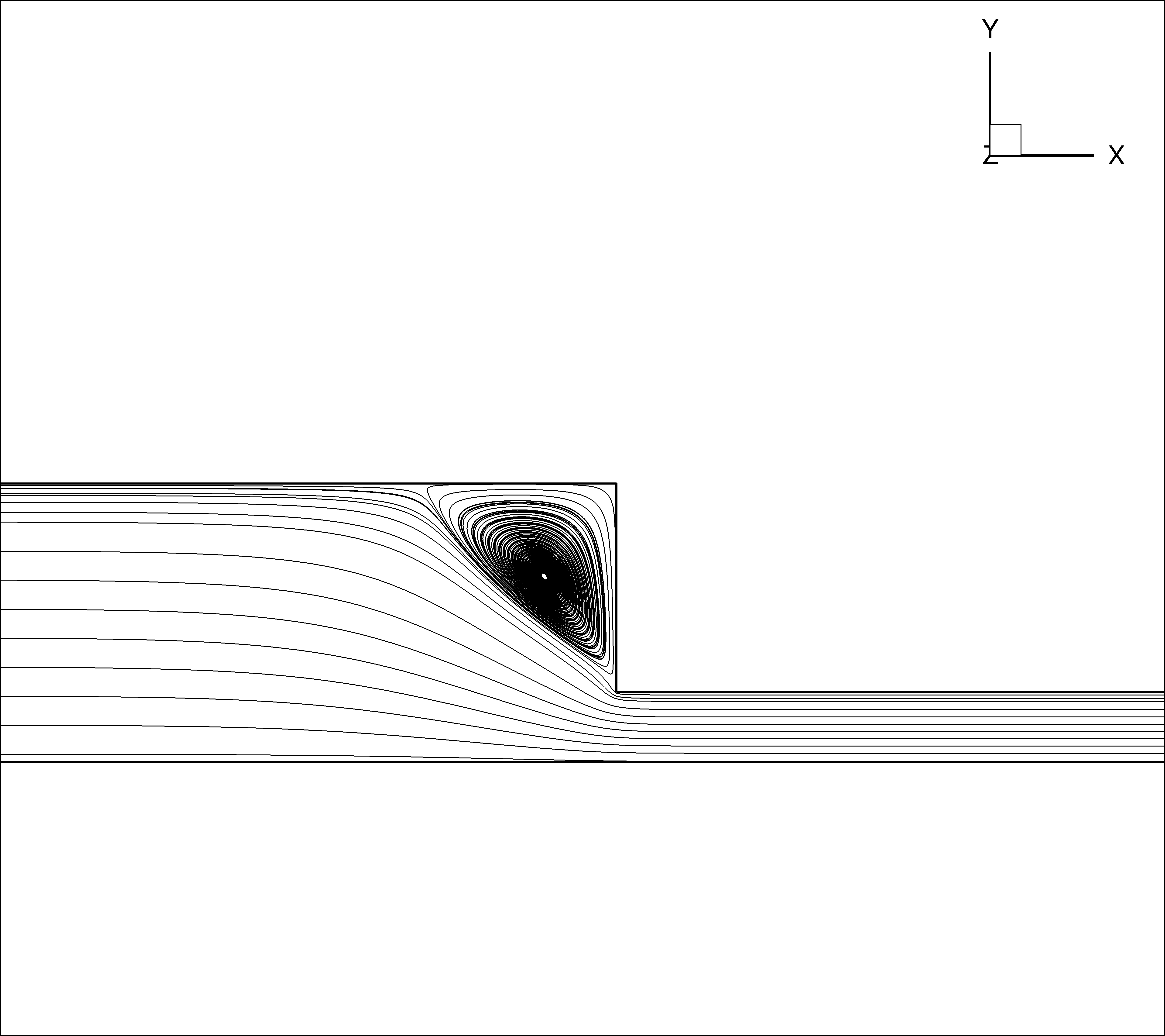}
	}%
\hspace{8pt}%
\subfigure[][]{%
 \label{fig:tw327b-d}%
\includegraphics[width=0.49\textwidth,trim=10mm 30cm 10mm 50cm,clip]{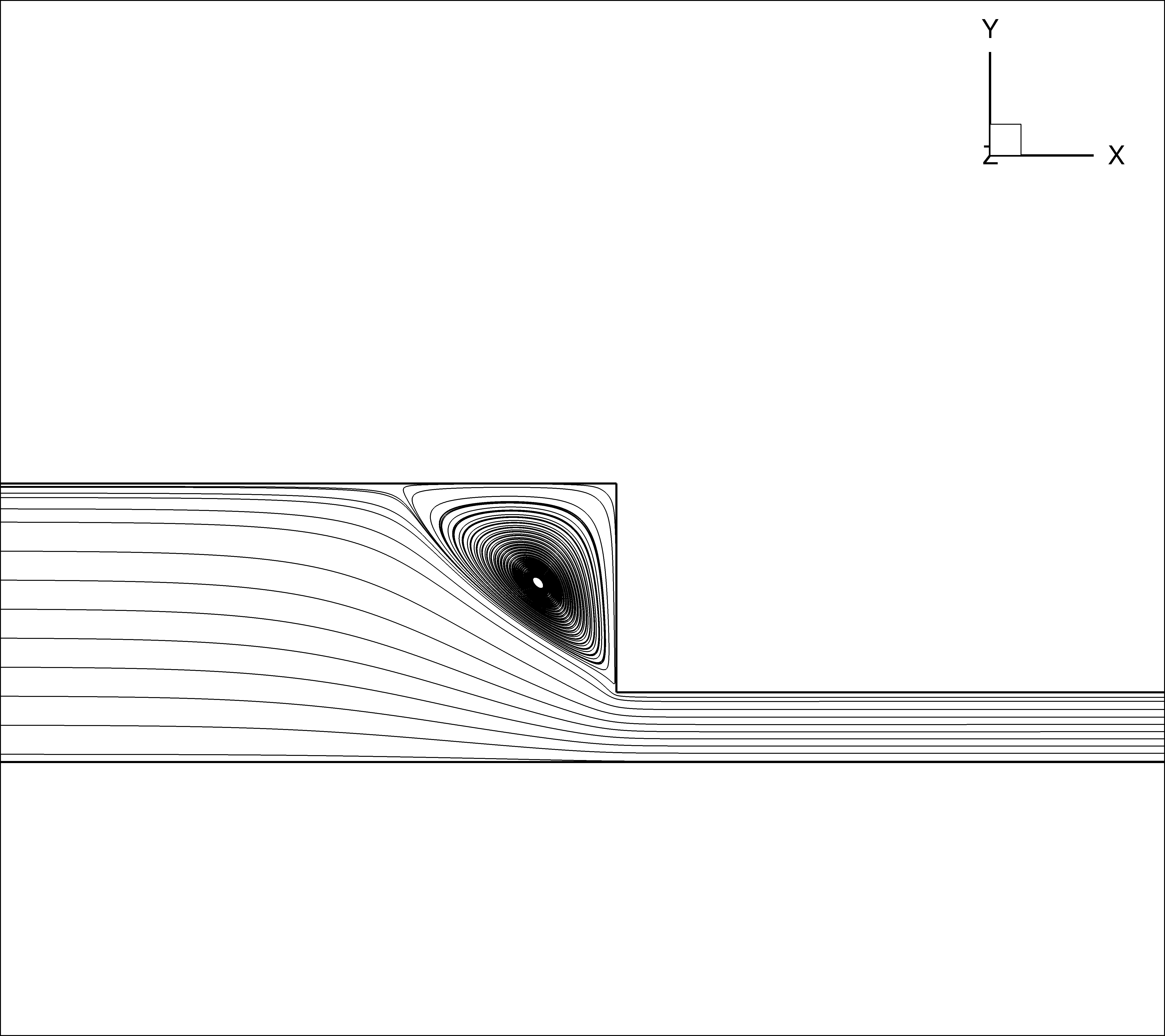}
  }\\
	\subfigure[][]{%
	 \label{fig:tw327b-e}
	\includegraphics[width=0.49\textwidth,trim=10mm 30cm 10mm 50cm,clip]{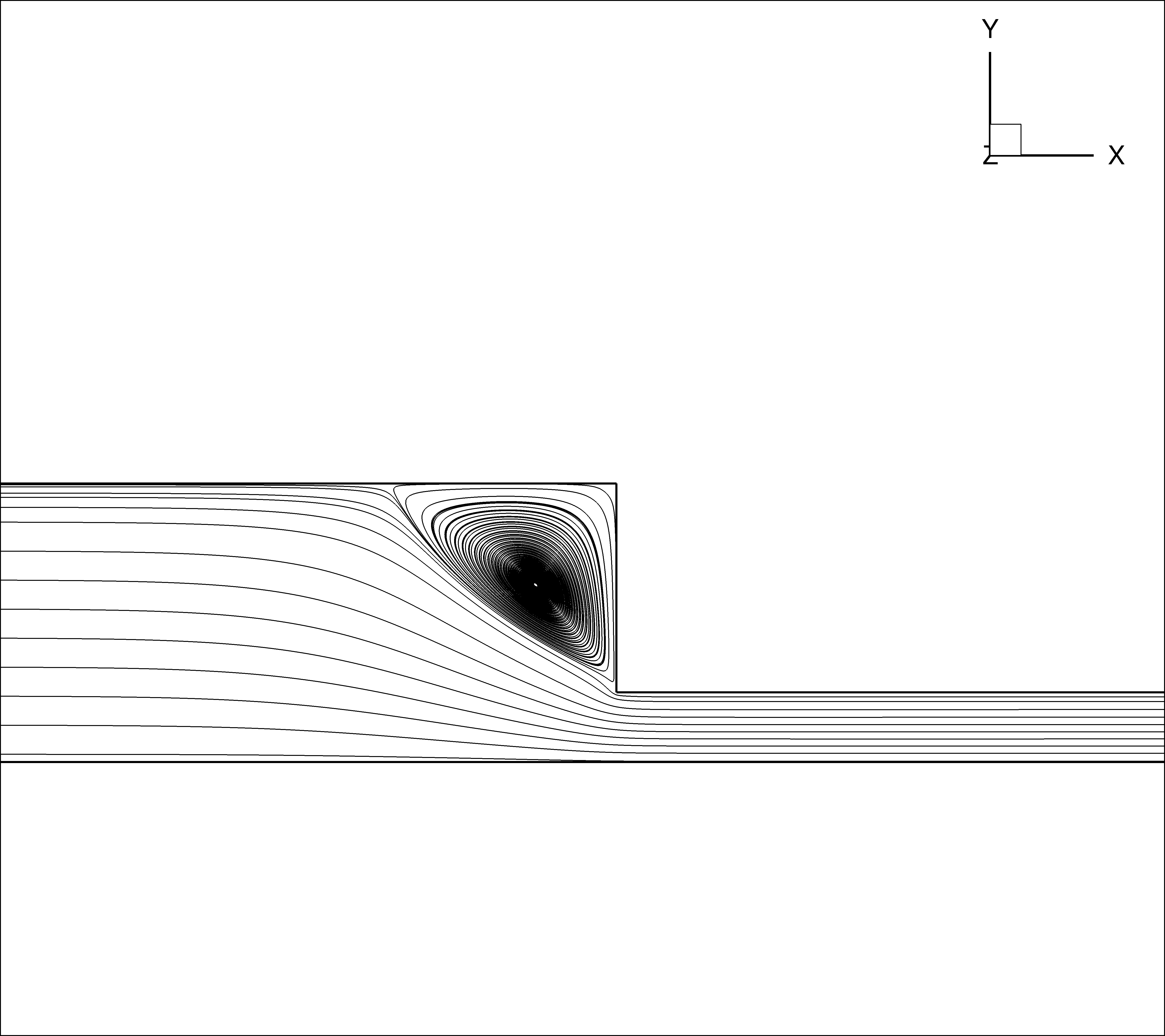}
   }%
\hspace{8pt}%
\subfigure[][]{%
	 \label{fig:tw327b-f}
	\includegraphics[width=0.49\textwidth,trim=10mm 30cm 10mm 50cm,clip]{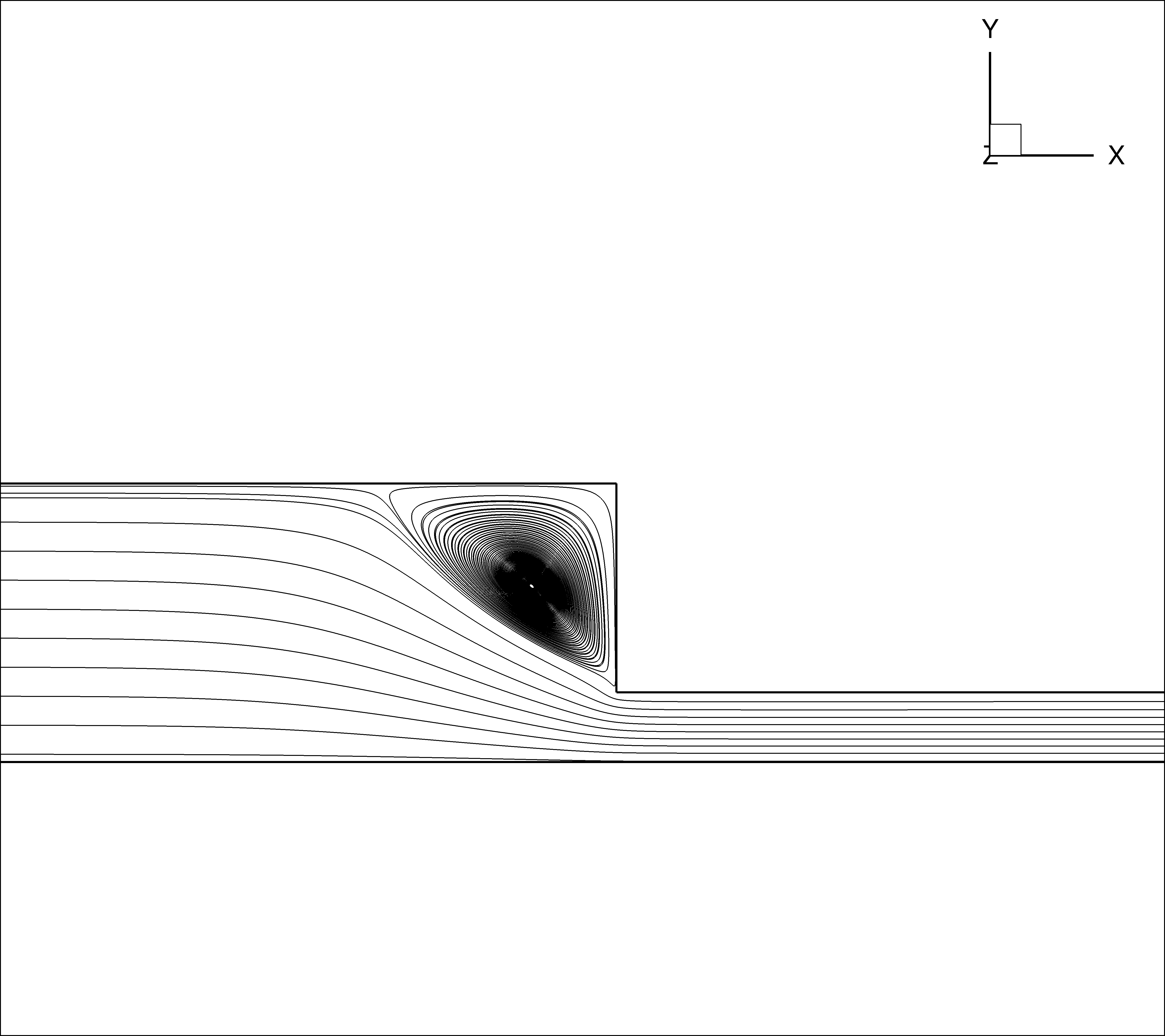}
}%
	\caption{Recirculation zone visualized by streamlines for
	\subref{fig:tw327b-a} $T_w=\SI{285}{K}$, $Wi=5$;
	\subref{fig:tw327b-b} $T_w=\SI{305}{K}$, $Wi=11.3$;
	\subref{fig:tw327b-c}-\subref{fig:tw327b-f} $T_w=\SI{327}{K}$ for Weissenberg numbers: 
	\subref{fig:tw327b-c} $Wi=4.0$
	\subref{fig:tw327b-d} $Wi=6.4$
	\subref{fig:tw327b-e} $Wi=8.2$ and
	\subref{fig:tw327b-f} $Wi=12.3$}
	\label{figure:tw327deb}%
\end{figure}

In Figures \ref{figure:tw327de_far} and \ref{figure:tw327de}, the development of temperature profiles with increasing Weissenberg number is displayed for wall temperature $T_{\text{w}} = \SI{327}{\K}$ at Weissenberg numbers $Wi=4.0, 6.4, 8.2$ and $12.3$. The simulations are performed on the finest mesh $3$.

Figure \ref{figure:tw327de_far} shows the temperature as a function of the radial position far upstream of the contraction plane at $\zeta=-8.0$. At this location, the temperature profile is assumed to be fully developed and not yet altered by the contraction.
We observe here, that the bulk temperature decreases with increasing Weissenberg number. As previously mentioned, the temperature change is underestimated in the simulation at all Weissenberg numbers. With increasing Weissenberg number, also in the experimental data $\theta$ approaches a value of one, which means $T \approx T_{\text{in}}$, in the center of the cylinder.
Here, the temperature is nearly unchanged compared to the inlet flow, which allows the conclusion that at high flow velocities only very little transport occurs in the radial direction. 
While the bulk temperature decreases with increasing Weissenberg number, the gradient of the temperature profile at the wall becomes steeper.

Figure \ref{figure:tw327de} shows temperature profiles near the contraction plane at $\zeta=-0.3$. 
At this position, we observe that with increasing Weissenberg number the bulk temperature decreases ($\theta$ at radial position $r/R_2=0$ approaches the value $1$). While the temperature is significantly underestimated in the simulations, this development can be observed both in experiments and simulation data. 
In the vicinity of the wall, a profile with a very low gradient is formed. The slope grows with increasing  Weissenberg numbers. At the lowest Weissenberg number, the temperature profile is almost linear,  whereas it becomes more curved at higher Weissenberg numbers.
While the values are slightly underestimated, this qualitative behavior is well captured by the simulation.
Interesting is the sharp bend in the temperature profiles, that can be observed both experimentally and numerically. It is slightly more pronounced in the simulations, especially at the highest Weissenberg number. At $Wi=12.3$, a considerably smoother temperature profile has been observed in the experimental setup, while in the simulation the sharp bend can still be found. Note that this observation of a smoother temperature profile at high Weissenberg number is not present in all test cases. Figures \ref{figure:tw305alphab}\subref{fig:tw305b-f} and \ref{figure:tw305alpha}\subref{fig:tw305a-f} show the temperature profiles for a nominal wall temperature $T_{\text{w}}=\SI{305}{\K}$ at Weissenberg numbers $Wi=11.3$ and $Wi=14.8$. In both cases, we observe a sharp bend of the temperature profile in the inner half of the cylinder both in experimental data and in simulation results.

Due to the quantitative deviations between the temperatures of the experiment and the simulation, the influence of the rheological model is investigated. Additional simulations are performed with the exponential PTT model, which better captures the extensional flow behavior close to the contraction. The results are shown in Figures \ref{figure:tw327de_far} and \ref{figure:tw327de}, where the dashed line represents the PTT model profiles with $\epsilon = 0.05$. We observe that a variation of the rheological model results in only small differences in the temperature profiles at the considered locations. Thus, the different stress distributions associated with the change of the rheological model have only a minor impact on the temperature in this setup. Besides the rheological model, variations of the thermal model might have a greater influence on the temperature prediction, but this goes beyond the scope of the present study.

Figure \ref{figure:tw327deb} visualizes the recirculation zone that forms in the upper corner of the contraction at the investigated wall temperatures and Weissenberg numbers for the Oldroyd-B fluid. We observe a growing length of the recirculation zone with increasing wall temperature. The recirculation zone is also growing with increasing Weissenberg numbers at the same wall temperature.

The presented results prove the stability of the suggested numerical framework at all investigated Weissenberg numbers. They show that the chosen thermo-rheological model is capable of achieving good qualitative agreement with the experimental data. Deviations between simulation and experimental data are found to become more pronounced at higher Weissenberg numbers, which is assumed to be due to the secondary flow in the experimental setup and the limitations of the chosen thermo-rheological model.

\subsubsection*{Splitting factor $\alpha$}
\begin{figure}%
\centering
\ref{dreinullfuenfalpha}\\
\subfigure[][]{%
  \label{fig:tw305b-a}%
	\begin{tikzpicture}
  \begin{axis}[
	legend style={legend columns=-1},
	legend to name=dreinullfuenfalpha,
	width=0.48\textwidth,
  xlabel={$r/R_2$},       
	ylabel={$\theta^{*}$},
	xlabel near ticks,
	ylabel near ticks,	
	ymin=0,
	ymax=1,
	xmin=0,
	xmax=4,
	minor y tick num=1,
	]
	\addplot [cyan,thick,dashed] table {Data/theta_exp03_dt30435_de113_al0_t800_far.csv};
	\addlegendentry{$\alpha=0$}
		\addplot [thick,dotted] table {Data/theta_exp03_dt30435_de113_al1_t800_far.csv};
	\addlegendentry{$\alpha=1$}
	\addplot [mark=diamond, only marks, black] table {Data/exp_yesi305_de113_far01.csv};
	\addlegendentry{experimental data \cite{yesilata2000}}
	\addplot [mark=diamond, only marks, black] table {Data/exp_yesi305_de113_far02.csv};
  \end{axis}
 \end{tikzpicture}}%
\hspace{8pt}%
\subfigure[][]{%
 \label{fig:tw305b-b}%
  \begin{tikzpicture}
  \begin{axis}[
	width=0.48\textwidth,
  xlabel={$r/R_2$},       
	ylabel={$\theta^{*}$},
	xlabel near ticks,
	ylabel near ticks,	
	ymin=0,
	ymax=1,
	xmin=0,
	xmax=4,
	minor y tick num=1,
	]	
	\addplot [cyan,thick,dashed] table {Data/theta_exp03_dt30435_de113_al0_t800_x06.csv};
	\addplot [thick,dotted] table {Data/theta_exp03_dt30435_de113_al1_t800_x06.csv};
  \end{axis}
 \end{tikzpicture}}\\
	\subfigure[][]{%
	 \label{fig:tw305b-c}
    \begin{tikzpicture}
  \begin{axis}[
	width=0.48\textwidth,
  xlabel={$r/R_2$},       
	ylabel={$\theta^{*}$},
	xlabel near ticks,
	ylabel near ticks,	
	ymin=0,
	ymax=1,
	xmin=0,
	xmax=4,
	minor y tick num=1,
	]
	\addplot [cyan,thick,dashed] table {Data/theta_exp03_dt30435_de113_al0_t800_x0602.csv};
	\addplot [thick,dotted] table {Data/theta_exp03_dt30435_de113_al1_t800_x0602.csv};
  \end{axis}
 \end{tikzpicture}}%
\hspace{8pt}%
\subfigure[][]{%
	 \label{fig:tw305b-d}
\begin{tikzpicture}
  \begin{axis}[
	width=0.48\textwidth,
  xlabel={$r/R_2$},       
	ylabel={$\theta^{*}$},
	xlabel near ticks,
	ylabel near ticks,	
	ymin=0,
	ymax=1,
	xmin=0,
	xmax=4,
	minor y tick num=1,
	]
	\addplot [cyan,thick,dashed] table {Data/theta_exp03_dt30435_de113_al0_t800_x0604.csv};
	\addplot [thick,dotted] table {Data/theta_exp03_dt30435_de113_al1_t800_x0604.csv};
  \end{axis}
 \end{tikzpicture}}\\
\subfigure[][]{%
 \label{fig:tw305b-e}%
  \begin{tikzpicture}
  \begin{axis}[
	width=0.48\textwidth,
  xlabel={$r/R_2$},       
	ylabel={$\theta^{*}$},
	xlabel near ticks,
	ylabel near ticks,	
	ymin=0,
	ymax=1,
	xmin=0,
	xmax=4,
	minor y tick num=1,
	]	
	\addplot [cyan,thick,dashed] table {Data/theta_exp03_dt30435_de113_al0_t800_x0606.csv};
	\addplot [thick,dotted] table {Data/theta_exp03_dt30435_de113_al1_t800_x0606.csv};
  \end{axis}
 \end{tikzpicture}}%
\hspace{8pt}%
	\subfigure[][]{%
	 \label{fig:tw305b-f}
    \begin{tikzpicture}
  \begin{axis}[
	width=0.48\textwidth,
  xlabel={$r/R_2$},       
	ylabel={$\theta^{*}$},
	xlabel near ticks,
	ylabel near ticks,	
	ymin=0,
	ymax=1,
	xmin=0,
	xmax=4,
	minor y tick num=1,
	]
	\addplot [cyan,thick,dashed] table {Data/theta_exp03_dt30435_de113_al0_t800.csv};
	\addplot [thick,dotted] table {Data/theta_exp03_dt30435_de113_al1_t800.csv};
	\addplot [mark=diamond, only marks, black] table {Data/exp_yesi305_de113_near_01.csv};
	\addplot [mark=diamond, only marks, black] table {Data/exp_yesi305_de113_near_02.csv};
  \end{axis}
 \end{tikzpicture}}%
	\caption{Dimensionless temperature $\theta^{*}$ vs. radial position at adjusted wall temperature $T_{\text{w}}=\SI{304.35}{K}$, $Wi=11.3$ at
	\subref{fig:tw305b-a}  $\zeta=-8.0$ 
	\subref{fig:tw305b-b}  $\zeta=-1.5$ 
	\subref{fig:tw305b-c}  $\zeta=-1.2$ 
	\subref{fig:tw305b-d}  $\zeta=-0.88$ 
	\subref{fig:tw305b-e}  $\zeta=-0.57$ and 
	\subref{fig:tw305b-f}  $\zeta=-0.3$} 
	\label{figure:tw305alphab}%
\end{figure}
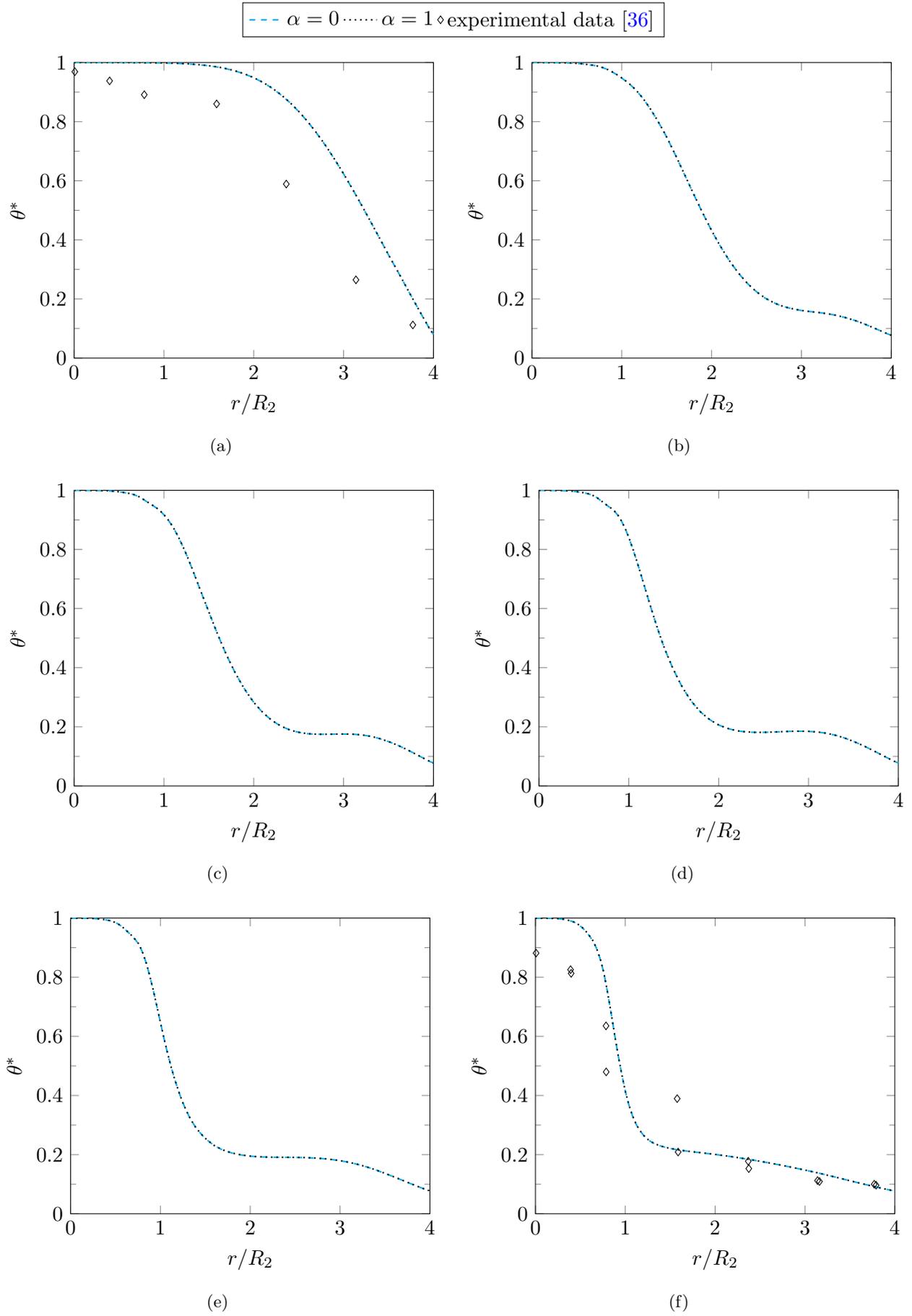
\begin{figure}%
\centering
\ref{dreinullfuenfalpha}\\
\subfigure[][]{%
  \label{fig:tw305a-a}%
	\begin{tikzpicture}
  \begin{axis}[
	width=0.48\textwidth,
  xlabel={$r/R_2$},       
	ylabel={$\theta$},
	xlabel near ticks,
	ylabel near ticks,	
	ymin=0,
	ymax=1,
	xmin=0,
	xmax=4,
	minor y tick num=1,
	]
	\addplot [cyan,thick,dashed] table {Data/theta_exp03_dt305_de148_al0_t600_far.csv};
	\addplot [thick,dotted] table {Data/theta_exp03_dt305_de148_al1_t600_far.csv};
	\addplot [mark=diamond, only marks, black] table {Data/exp_yesi305_de148_far01.csv};
	\addplot [mark=diamond, only marks, black] table {Data/exp_yesi305_de148_far02.csv};
  \end{axis}
 \end{tikzpicture}}%
\hspace{8pt}%
\subfigure[][]{%
 \label{fig:tw305a-b}%
  \begin{tikzpicture}
  \begin{axis}[
	width=0.48\textwidth,
  xlabel={$r/R_2$},       
	ylabel={$\theta$},
	xlabel near ticks,
	ylabel near ticks,	
	ymin=0,
	ymax=1,
	xmin=0,
	xmax=4,
	minor y tick num=1,
	]	
	\addplot [cyan,thick,dashed] table {Data/theta_exp03_dt305_de148_al0_t600_x6.csv};
	\addplot [thick,dotted] table {Data/theta_exp03_dt305_de148_al1_t600_x6.csv};
  \end{axis}
 \end{tikzpicture}}\\
	\subfigure[][]{%
	 \label{fig:tw305a-c}
    \begin{tikzpicture}
  \begin{axis}[
	width=0.48\textwidth,
  xlabel={$r/R_2$},       
	ylabel={$\theta$},
	xlabel near ticks,
	ylabel near ticks,	
	ymin=0,
	ymax=1,
	xmin=0,
	xmax=4,
	minor y tick num=1,
	]
	\addplot [cyan,thick,dashed] table {Data/theta_exp03_dt305_de148_al0_t600_x602.csv};
	\addplot [thick,dotted] table {Data/theta_exp03_dt305_de148_al1_t600_x602.csv};
  \end{axis}
 \end{tikzpicture}}%
\hspace{8pt}%
\subfigure[][]{%
	 \label{fig:tw305a-d}
\begin{tikzpicture}
  \begin{axis}[
	width=0.48\textwidth,
  xlabel={$r/R_2$},       
	ylabel={$\theta$},
	xlabel near ticks,
	ylabel near ticks,	
	ymin=0,
	ymax=1,
	xmin=0,
	xmax=4,
	minor y tick num=1,
	]
	\addplot [cyan,thick,dashed] table {Data/theta_exp03_dt305_de148_al0_t600_x604.csv};
	\addplot [thick,dotted] table {Data/theta_exp03_dt305_de148_al1_t600_x604.csv};
  \end{axis}
 \end{tikzpicture}}\\
\subfigure[][]{%
 \label{fig:tw305a-e}%
  \begin{tikzpicture}
  \begin{axis}[
	width=0.48\textwidth,
  xlabel={$r/R_2$},       
	ylabel={$\theta$},
	xlabel near ticks,
	ylabel near ticks,	
	ymin=0,
	ymax=1,
	xmin=0,
	xmax=4,
	minor y tick num=1,
	]	
	\addplot [cyan,thick,dashed] table {Data/theta_exp03_dt305_de148_al0_t600_x606.csv};
	\addplot [thick,dotted] table {Data/theta_exp03_dt305_de148_al1_t600_x606.csv};
  \end{axis}
 \end{tikzpicture}}%
\hspace{8pt}%
	\subfigure[][]{%
	 \label{fig:tw305a-f}
    \begin{tikzpicture}
  \begin{axis}[
	width=0.48\textwidth,
  xlabel={$r/R_2$},       
	ylabel={$\theta$},
	xlabel near ticks,
	ylabel near ticks,	
	ymin=0,
	ymax=1,
	xmin=0,
	xmax=4,
	minor y tick num=1,
	]
	\addplot [cyan,thick,dashed] table {Data/theta_exp03_dt305_de148_al0_t600.csv};
	\addplot [thick,dotted] table {Data/theta_exp03_dt305_de148_al1_t600.csv};	
	\addplot [mark=diamond, only marks, black] table {Data/exp_yesi305_de148_near01.csv};
	\addplot [mark=diamond, only marks, black] table {Data/exp_yesi305_de148_near02.csv};
  \end{axis}
 \end{tikzpicture}}%
	\caption{Dimensionless temperature vs. radial position at $T_{\text{w}}=\SI{305}{\K}$, $Wi=14.8$ at
	\subref{fig:tw305a-a} $\zeta=-8.0$
	\subref{fig:tw305a-b} $\zeta=-1.5$
	\subref{fig:tw305a-c} $\zeta=-1.2$ 
	\subref{fig:tw305a-d} $\zeta=-0.88$
	\subref{fig:tw305a-e} $\zeta=-0.57$ and
	\subref{fig:tw305a-f} $\zeta=-0.3$}
	\label{figure:tw305alpha}%
\end{figure}
  	
\begin{figure}%
\centering
\ref{dreinullfuenfbeta}\\
\subfigure[][]{%
  \label{fig:tw327a-a}%
	\begin{tikzpicture}
  \begin{axis}[
	legend style={legend columns=-1},
	legend to name=dreinullfuenfbeta,
	width=0.48\textwidth,
  xlabel={$r/R_2$},       
	ylabel={$\theta$},
	xlabel near ticks,
	ylabel near ticks,	
	ymin=0,
	ymax=1,
	xmin=0,
	xmax=4,
	minor y tick num=1,
	]
	\addplot [cyan,thick,dashed] table {Data/theta_exp03_dt327_de148_al0_t400_far.csv};
		\addlegendentry{$\alpha=0$}
	\addplot [thick,dotted] table {Data/theta_exp03_dt327_de148_al1_t400_far.csv};
		\addlegendentry{$\alpha=1$}
  \end{axis}
 \end{tikzpicture}}%
\hspace{8pt}%
\subfigure[][]{%
 \label{fig:tw327a-b}%
  \begin{tikzpicture}
  \begin{axis}[
	width=0.48\textwidth,
  xlabel={$r/R_2$},       
	ylabel={$\theta$},
	xlabel near ticks,
	ylabel near ticks,	
	ymin=0,
	ymax=1,
	xmin=0,
	xmax=4,
	minor y tick num=1,
	]	
	\addplot [cyan,thick,dashed] table {Data/theta_exp03_dt327_de148_al0_t400_x06.csv};
	\addplot [thick,dotted] table {Data/theta_exp03_dt327_de148_al1_t400_x06.csv};
  \end{axis}
 \end{tikzpicture}}\\
	\subfigure[][]{%
	 \label{fig:tw327a-c}
    \begin{tikzpicture}
  \begin{axis}[
	width=0.48\textwidth,
  xlabel={$r/R_2$},       
	ylabel={$\theta$},
	xlabel near ticks,
	ylabel near ticks,	
	ymin=0,
	ymax=1,
	xmin=0,
	xmax=4,
	minor y tick num=1,
	]
	\addplot [cyan,thick,dashed] table {Data/theta_exp03_dt327_de148_al0_t400_x0602.csv};
	\addplot [thick,dotted] table {Data/theta_exp03_dt327_de148_al1_t400_x0602.csv};
  \end{axis}
 \end{tikzpicture}}%
\hspace{8pt}%
\subfigure[][]{%
	 \label{fig:tw327a-d}
\begin{tikzpicture}
  \begin{axis}[
	width=0.48\textwidth,
  xlabel={$r/R_2$},       
	ylabel={$\theta$},
	xlabel near ticks,
	ylabel near ticks,	
	ymin=0,
	ymax=1,
	xmin=0,
	xmax=4,
	minor y tick num=1,
	]
	\addplot [cyan,thick,dashed] table {Data/theta_exp03_dt327_de148_al0_t400_x0604.csv};
	\addplot [thick,dotted] table {Data/theta_exp03_dt327_de148_al1_t400_x0604.csv};
  \end{axis}
 \end{tikzpicture}}\\
\subfigure[][]{%
 \label{fig:tw327a-e}%
  \begin{tikzpicture}
  \begin{axis}[
	width=0.48\textwidth,
  xlabel={$r/R_2$},       
	ylabel={$\theta$},
	xlabel near ticks,
	ylabel near ticks,	
	ymin=0,
	ymax=1,
	xmin=0,
	xmax=4,
	minor y tick num=1,
	]	
	\addplot [cyan,thick,dashed] table {Data/theta_exp03_dt327_de148_al0_t400_x0606.csv};
	\addplot [thick,dotted] table {Data/theta_exp03_dt327_de148_al1_t400_x0606.csv};
  \end{axis}
 \end{tikzpicture}}%
\hspace{8pt}%
	\subfigure[][]{%
	 \label{fig:tw327a-f}
    \begin{tikzpicture}
  \begin{axis}[
	width=0.48\textwidth,
  xlabel={$r/R_2$},       
	ylabel={$\theta$},
	xlabel near ticks,
	ylabel near ticks,	
	ymin=0,
	ymax=1,
	xmin=0,
	xmax=4,
	minor y tick num=1,
	]
	\addplot [cyan,thick,dashed] table {Data/theta_exp03_dt327_de148_al0_t400.csv};
	\addplot [thick,dotted] table {Data/theta_exp03_dt327_de148_al1_t400.csv};	
  \end{axis}
 \end{tikzpicture}}%
	\caption{Dimensionless temperature vs. radial position at $T_{\text{w}}=\SI{327}{\K}$, $Wi=14.8$ at
	\subref{fig:tw327a-a} $\zeta=-8.0$
	\subref{fig:tw327a-b} $\zeta=-1.5$
	\subref{fig:tw327a-c} $\zeta=-1.2$ 
	\subref{fig:tw327a-d} $\zeta=-0.88$
	\subref{fig:tw327a-e} $\zeta=-0.57$ and
	\subref{fig:tw327a-f} $\zeta=-0.3$}
	\label{figure:tw327alpha}%
\end{figure}

\begin{figure}%
\centering
\ref{dreinullfuenfbeta}\\
\subfigure[][]{%
  \label{fig:tw285a-a}%
	\begin{tikzpicture}
  \begin{axis}[
	width=0.48\textwidth,
  xlabel={$r/R_2$},       
	ylabel={$\theta$},
	xlabel near ticks,
	ylabel near ticks,	
	ymin=0,
	ymax=1,
	xmin=0,
	xmax=4,
	minor y tick num=1,
	]
	\addplot [cyan,thick,dashed] table {Data/theta_exp03_dt285_de148_al0_t400_far.csv};
	\addplot [thick,dotted] table {Data/theta_exp03_dt285_de148_al1_t400_far.csv};
  \end{axis}
 \end{tikzpicture}}%
\hspace{8pt}%
\subfigure[][]{%
 \label{fig:tw285a-b}%
  \begin{tikzpicture}
  \begin{axis}[
	width=0.48\textwidth,
  xlabel={$r/R_2$},       
	ylabel={$\theta$},
	xlabel near ticks,
	ylabel near ticks,	
	ymin=0,
	ymax=1,
	xmin=0,
	xmax=4,
	minor y tick num=1,
	]	
	\addplot [cyan,thick,dashed] table {Data/theta_exp03_dt285_de148_al0_t400_x06.csv};
	\addplot [thick,dotted] table {Data/theta_exp03_dt285_de148_al1_t400_x06.csv};
  \end{axis}
 \end{tikzpicture}}\\
	\subfigure[][]{%
	 \label{fig:tw285a-c}
    \begin{tikzpicture}
  \begin{axis}[
	width=0.48\textwidth,
  xlabel={$r/R_2$},       
	ylabel={$\theta$},
	xlabel near ticks,
	ylabel near ticks,	
	ymin=0,
	ymax=1,
	xmin=0,
	xmax=4,
	minor y tick num=1,
	]
	\addplot [cyan,thick,dashed] table {Data/theta_exp03_dt285_de148_al0_t400_x0602.csv};
	\addplot [thick,dotted] table {Data/theta_exp03_dt285_de148_al1_t400_x0602.csv};
  \end{axis}
 \end{tikzpicture}}%
\hspace{8pt}%
\subfigure[][]{%
	 \label{fig:tw285a-d}
\begin{tikzpicture}
  \begin{axis}[
	width=0.48\textwidth,
  xlabel={$r/R_2$},       
	ylabel={$\theta$},
	xlabel near ticks,
	ylabel near ticks,	
	ymin=0,
	ymax=1,
	xmin=0,
	xmax=4,
	minor y tick num=1,
	]
	\addplot [cyan,thick,dashed] table {Data/theta_exp03_dt285_de148_al0_t400_x0604.csv};
	\addplot [thick,dotted] table {Data/theta_exp03_dt285_de148_al1_t400_x0604.csv};
  \end{axis}
 \end{tikzpicture}}\\
\subfigure[][]{%
 \label{fig:tw285a-e}%
  \begin{tikzpicture}
  \begin{axis}[
	width=0.48\textwidth,
  xlabel={$r/R_2$},       
	ylabel={$\theta$},
	xlabel near ticks,
	ylabel near ticks,	
	ymin=0,
	ymax=1,
	xmin=0,
	xmax=4,
	minor y tick num=1,
	]	
	\addplot [cyan,thick,dashed] table {Data/theta_exp03_dt285_de148_al0_t400_x0606.csv};
	\addplot [thick,dotted] table {Data/theta_exp03_dt285_de148_al1_t400_x0606.csv};
  \end{axis}
 \end{tikzpicture}}%
\hspace{8pt}%
	\subfigure[][]{%
	 \label{fig:tw285a-f}
    \begin{tikzpicture}
  \begin{axis}[
	width=0.48\textwidth,
  xlabel={$r/R_2$},       
	ylabel={$\theta$},
	xlabel near ticks,
	ylabel near ticks,	
	ymin=0,
	ymax=1,
	xmin=0,
	xmax=4,
	minor y tick num=1,
	]
	\addplot [cyan,thick,dashed] table {Data/theta_exp03_dt285_de148_al0_t400.csv};
	\addplot [thick,dotted] table {Data/theta_exp03_dt285_de148_al1_t400.csv};	
  \end{axis}
 \end{tikzpicture}}%
	\caption{Dimensionless temperature vs. radial position at $T_{\text{w}}=\SI{285}{\K}$, $Wi=14.8$ at
	\subref{fig:tw285a-a} $\zeta=-8.0$
	\subref{fig:tw285a-b} $\zeta=-1.5$
	\subref{fig:tw285a-c} $\zeta=-1.2$ 
	\subref{fig:tw285a-d} $\zeta=-0.88$
	\subref{fig:tw285a-e} $\zeta=-0.57$ and
	\subref{fig:tw285a-f} $\zeta=-0.3$}
	\label{figure:tw285alpha}%
\end{figure}

All calculations shown so far were calculated with an arbitrarily chosen value of the splitting factor of $\alpha=0.5$. 
For an estimation of the influence of the splitting factor, calculations with the two limiting cases of pure energy elasticity ($\alpha=0$) and pure entropy elasticity ($\alpha=1$) were performed at the highest investigated Weissenberg number $Wi=14.8$ and at $Wi=11.3$ with the same nominal wall temperature for comparison.

In Figure \ref{figure:tw305alphab}, simulation results at Weissenberg number $Wi=11.3$ with an adjusted wall temperature of $T_{\text{w}}=\SI{304.35}{\K}$ are shown at various axial positions. The dashed line represents the calculation with $\alpha=0$, the dotted line represents the calculation with $\alpha=1$. Figure~\ref{figure:tw305alphab}$(a)$ corresponds to the dimensionless position $\zeta=-8$ and Fig.~\ref{figure:tw305alphab}$(f)$ to position $\zeta=-0.3$, at which the temperature was measured in the cited experiments. We observe no difference in the dimensionless temperature profiles between the limiting cases of $\alpha=0$ and $\alpha=1$.
Figure \ref{figure:tw305alpha} presents calculations for a nominal wall temperature of $T_{\text{w}}=\SI{305}{\K}$ and Weissenberg number $Wi=14.8$ at various axial positions.

At this Weissenberg number, we find no temperature difference at $\zeta=-8$, a location far upstream of the contraction and upstream of the recirculation zone. Visible deviations are present for all consecutive temperature profiles that are located inside the recirculation zone. At $\alpha=1$, the $\theta$ values are lower, so the temperature is higher than for $\alpha=0$. This is in accordance with our expectations: at $\alpha=1$ all mechanical energy is dissipated resulting in a higher temperature rise. In case of $\alpha=0$, part of the mechanical energy is stored and the temperature rise is less significant. We find that the difference gets smaller when approaching the contraction.

Peters et al.~\cite{peters1997} assumed that the difference in temperature between pure energy and pure entropy elasticity increases with increasing Weissenberg number. Our observations lead to the same conclusion. While the deviation in temperature between the limiting cases of $\alpha=0$ and $\alpha=1$ is negligible at Weissenberg number $Wi=11.3$, it is obviously present at Weissenberg number $Wi=14.8$ for an imposed wall temperature of $T_{\text{w}}=\SI{305}{\K}$. This suggests that the importance of energy storage is not yet very pronounced at lower Weissenberg numbers. With regard to the results presented so far in this section, an arbitrary choice of $\alpha$ in the limits $0 \leq \alpha \leq 1$ seems justified at small Weissenberg numbers for the considered test case. Experimental probe data at additional positions inside the recirculation zone would be necessary to estimate a fitting value of $\alpha$ for a specific test case.

Finally, we investigate how the imposed wall temperature affects energy partitioning. 
Figures \ref{figure:tw327alpha} and \ref{figure:tw285alpha} show temperature profiles of simulations at Weissenberg number $Wi=14.8$ and wall temperatures $T_{\text{w}}=\SI{327}{\K}$ and $T_{\text{w}}=\SI{285}{\K}$, respectively. In the case of a heated wall, the same observations are valid as for the above-described wall temperature of $T_{\text{w}}=\SI{305}{\K}$ at the same Weissenberg number. Deviations between the calculations at $\alpha=0$ and $\alpha=1$ are present inside the recirculation zone and decrease when approaching the contraction. Regarding the cooled wall, we find no difference in the temperature profiles for pure energy elasticity and pure entropy elasticity. It is possible that in this case, the importance of the energy partitioning for the flow field starts at higher Weissenberg numbers. We conclude that, apart from the Weissenberg number, also the imposed wall temperature affects the division between energy and entropy elasticity.

\section{Summary and conclusions}
An original approach for modeling the non-isothermal flow of Oldroyd-B type fluids at high Weissenberg numbers is developed. The implementation is based on an established FV framework for isothermal viscoelastic flows~\cite{niethammer2018, niethammer2019, niethammer2019b, niethammer2019c}.
Stable calculations at high Weissenberg numbers are ensured by the root conformation representation, which is extended to non-isothermal flows in this study.
The temperature dependence of the constitutive equation is modeled by the time-temperature superposition principle. For the internal energy equation, Fourier's law is used for heat conduction, and partitioning between energy and entropy elasticity is realized with a constant splitting factor.

An analytical solution of the Oldroyd-B fluid in planar channel flow is derived for the field variables velocity, first normal stress and temperature. The temperature dependence of viscosity and relaxation time are neglected in these results. The simulation data are compared to the analytical flow profiles to verify the solution and to estimate the numerical error. All considered field variables show good agreement with the analytical data and the error is found to decrease quadratically with mesh refinement.

The validation of the solver is performed with experimental data from \cite{yesilata2000}, where a highly elastic polyisobutylene-based polymer solution is investigated in a circular 4:1 contraction. We perform simulations in an axisymmetric setup that is modeled as close as possible to the experimental geometry.
Profiles of dimensionless temperature over radial position are compared at different wall temperatures.
The results indicate a good qualitative reproduction of the measured temperature profiles. 
The deviations are most pronounced in the middle of the cylinder. 
The change of the bulk temperature is more significant in the experimental data, while 
we observe only small changes of the bulk temperature in the numerical simulations.
Since heat conduction is comparably low in viscoelastic fluids, the main driving force for these temperature changes is heat production by viscous dissipation. 
In any natural flow we expect asymmetric flow profiles, including secondary flow that contains velocity gradients and is an additional source of viscous dissipation. The random asymmetry is not modeled in the numerical setup. As a consequence, the numerical solution with symmetric flow profiles could be regarded as a flow at ideal conditions, with minimal viscous dissipation. The results indicate that less thermal energy is converted in the simulation and it can thus be regarded as a lower bound for converted energy.

Comparative computations with the exponential PTT model show that a variation of the rheological model has only minor influence on the temperature field at the considered locations.

The comparison of the simulation results at Weissenberg numbers between $Wi=5$ and $Wi=14.8$ to experimental data shows that the chosen thermo-rheological model is capable of achieving good qualitative agreement with the experimental data. The results prove the stability of the suggested numerical framework at all investigated Weissenberg numbers. Deviations between simulation and experimental data are found to become more pronounced at higher Weissenberg numbers, which is assumed to be due to secondary flow in the experimental setup and the limitations of the chosen thermo-rheological model.

The energy partitioning factor $\alpha$ is a purely modeling constant. We find that an arbitrary choice in the range of $0 \leq \alpha \leq 1$ is justified at low Weissenberg numbers, as the deviation in the solutions of the two limiting cases of pure entropy elasticity and pure energy elasticity are negligible. For $Wi=14.8$ and heated walls, we find significant deviations in temperature inside the recirculation zone, confirming the assumption by Peters and Baaijens \cite{peters1997} that the effect of the energy partitioning becomes more important at higher Weissenberg numbers. 
We also find a dependence of the energy partitioning on the imposed wall temperature. 
For cooled walls, we observe no deviation at the same Weissenberg number, leading to the possible conclusion that in these flow regimes the Weissenberg number at which energy partitioning becomes important for the flow field is significantly higher than in the case of heated walls. 
 
\section{Acknowledgments}
The work of the first author is supported by the Graduate School CE within the Centre for Computational Engineering at 
Technische Universit{\"a}t Darmstadt. Calculations for this research were conducted on the Lichtenberg high performance computer of the TU Darmstadt.

\bibliographystyle{abbrv}
\bibliography{literature_paper2019}

\begin{thebibliography}{10}

\bibitem{balci2011}
N.~Balci, B.~Thomases, M.~Renardy, and C.~R. Doering.
\newblock {Symmetric factorization of the conformation tensor in viscoelastic
  fluid models}.
\newblock {\em J. Non-Newton. Fluid Mech.}, 166(11):546--553, 2011.

\bibitem{bird1995}
R.~B. Bird.
\newblock {Constitutive Equations for Polymeric Liquids}.
\newblock {\em Annu. Rev. Fluid Mech.}, 27(1):169--193, 1995.

\bibitem{bird1987dynamics}
R.~B. Bird, R.~C. Armstrong, and O.~Hassager.
\newblock {\em Dynamics of polymeric liquids. Vol. 1: Fluid mechanics}.
\newblock John Wiley and Sons Inc., New York, NY, 1987.

\bibitem{boger1985}
D.~V. Boger.
\newblock {Model polymer fluid systems}.
\newblock {\em Pure Appl. Chem.}, 57(7):921--930, 1985.

\bibitem{boger1987}
D.~V. Boger.
\newblock {Viscoelastic Flows Through Contractions}.
\newblock {\em Annu. Rev. Fluid Mech.}, 19:157--182, 1987.

\bibitem{bowdler1971}
H.~Bowdler, R.~S. Martin, C.~Reinsch, and J.~H. Wilkinson.
\newblock {The QR and QL Algorithms for Symmetric Matrices}.
\newblock In {\em Handb. Autom. Comput.}, pages 227--240. Springer Berlin
  Heidelberg, 1971.

\bibitem{braun1991}
H.~Braun.
\newblock {A model for the thermorheological behavior of viscoelastic fluids}.
\newblock {\em Rheol. Acta}, 30(6):523--529, 1991.

\bibitem{braun1990}
H.~Braun and C.~Friedrich.
\newblock {Dissipative behaviour of viscoelastic fluids derived from
  rheological constitutive equations}.
\newblock {\em J. Non-Newton. Fluid Mech.}, 38(1):81--91, 1990.

\bibitem{fattal2004}
R.~Fattal and R.~Kupferman.
\newblock {Constitutive laws for the matrix-logarithm of the conformation
  tensor}.
\newblock {\em J. Non-Newton. Fluid Mech.}, 123(2-3):281--285, 2004.

\bibitem{fattal2005}
R.~Fattal and R.~Kupferman.
\newblock {Time-dependent simulation of viscoelastic flows at high Weissenberg
  number using the log-conformation representation}.
\newblock {\em J. Non-Newton. Fluid Mech.}, 126(1):23--37, 2005.

\bibitem{ferry1980}
J.~D. Ferry.
\newblock {\em Viscoelastic properties of polymers}.
\newblock John Wiley \& Sons, New York, NY, 1980.

\bibitem{guenette1995}
R.~Gu{\'{e}}nette and M.~Fortin.
\newblock {A new mixed finite element method for computing viscoelastic flows}.
\newblock {\em J. Non-Newton. Fluid Mech.}, 60(1):27--52, 1995.

\bibitem{habla2012}
F.~Habla, A.~Woitalka, S.~Neuner, and O.~Hinrichsen.
\newblock {Development of a methodology for numerical simulation of
  non-isothermal viscoelastic fluid flows with application to axisymmetric 4:1
  contraction flows}.
\newblock {\em Chem. Eng. J.}, pages 772--784, 2012.

\bibitem{huetter2009}
M.~H{\"{u}}tter, C.~Luap, and H.~C. {\"{O}}ttinger.
\newblock {Energy elastic effects and the concept of temperature in flowing
  polymeric liquids}.
\newblock {\em Rheol. Acta}, 48(3):301--316, 2009.

\bibitem{joseph1985}
D.~D. Joseph, M.~Renardy, and J.-C. Saut.
\newblock {Hyperbolicity and change of type in the flow of viscoelastic
  fluids}.
\newblock {\em Arch. Ration. Mech. Anal.}, 87(3):213--251, 1985.

\bibitem{keunings1986}
R.~Keunings.
\newblock {On the high Weissenberg number problem}.
\newblock {\em J. Non-Newton. Fluid Mech.}, 20:209--226, 1986.

\bibitem{darwish2015}
F.~Moukalled, L.~Mangani, and M.~Darwish.
\newblock {\em {The Finite Volume Method in Computational Fluid Dynamics}}.
\newblock Fluid Mechanics and Its Applications. Springer International
  Publishing, 2016.

\bibitem{niethammer2019c}
M.~Niethammer.
\newblock {\em {A Finite Volume Framework for Viscoelastic Flows at High
  Weissenberg Number}}.
\newblock PhD thesis, Technische Universit{\"{a}}t Darmstadt, 2019.

\bibitem{niethammer2019b}
M.~Niethammer, G.~Brenn, H.~Marschall, and D.~Bothe.
\newblock {An extended volume of fluid method and its application to single
  bubbles rising in a viscoelastic liquid}.
\newblock {\em J. Comput. Phys.}, 387:326--355, 2019.

\bibitem{niethammer2019}
M.~Niethammer, H.~Marschall, and D.~Bothe.
\newblock {Robust Direct Numerical Simulation of Viscoelastic Flows}.
\newblock {\em Chemie Ing. Tech.}, 91(4):522--528, 2019.

\bibitem{niethammer2018}
M.~Niethammer, H.~Marschall, C.~Kunkelmann, and D.~Bothe.
\newblock {A numerical stabilization framework for viscoelastic fluid flow
  using the finite volume method on general unstructured meshes}.
\newblock {\em Int. J. Numer. Methods Fluids}, 86(2):131--166, 2018.

\bibitem{peters1997}
G.~W. Peters and F.~P. Baaijens.
\newblock {Modelling of non-isothermal viscoelastic flows}.
\newblock {\em J. Non-Newton. Fluid Mech.}, 68(2-3):205--224, 1997.

\bibitem{phanthien1977}
N.~Phan~Thien and R.~I. Tanner.
\newblock A new constitutive equation derived from network theory.
\newblock {\em J. Non-Newton. Fluid Mech.}, 2(4):353 -- 365, 1977.

\bibitem{rhiechow1983}
C.~M. Rhie and W.~L. Chow.
\newblock {Numerical study of the turbulent flow past an airfoil with trailing
  edge separation}.
\newblock {\em AIAA J.}, 21(11):1525--1532, 1983.

\bibitem{shaw2012}
M.~T. Shaw.
\newblock {\em {Introduction to Polymer Rheology}}.
\newblock John Wiley {\&} Sons, Inc., Hoboken, NJ, USA, 2011.

\bibitem{spanjaards2020}
M.~Spanjaards, M.~Hulsen, and P.~Anderson.
\newblock Computational analysis of the extrudate shape of three-dimensional
  viscoelastic, non-isothermal extrusion flows.
\newblock {\em J. Non-Newton. Fluid Mech.}, page 104310, 2020.

\bibitem{sweby}
P.~K. Sweby.
\newblock {High Resolution Schemes Using Flux Limiters for Hyperbolic
  Conservation Laws}.
\newblock {\em SIAM J. Numer. Anal.}, 21(5):995--1011, 1984.

\bibitem{vanLeer1974}
B.~van Leer.
\newblock {Towards the ultimate conservative difference scheme. II.
  Monotonicity and conservation combined in a second-order scheme}.
\newblock {\em J. Comput. Phys.}, 14(4):361--370, 1974.

\bibitem{wachs2000}
A.~Wachs and J.-R. Clermont.
\newblock {Non-isothermal viscoelastic flow computations in an axisymmetric
  contraction at high Weissenberg numbers by a finite volume method}.
\newblock {\em J. Non-Newton. Fluid Mech.}, 95(2-3):147--184, 2000.

\bibitem{wachs2002}
A.~Wachs, J.-R. Clermont, and A.~Khalifeh.
\newblock Computations of non-isothermal viscous and viscoelastic flows in
  abrupt contractions using a finite volume method.
\newblock {\em Engineering Computations}, 19(8):874--901, 2002.

\bibitem{wapperom1998thermo}
P.~Wapperom and M.~A. Hulsen.
\newblock {Thermodynamics of viscoelastic fluids: The temperature equation}.
\newblock {\em J. Rheol.}, 42(5):999--1019, 1998.

\bibitem{weller1998}
H.~G. Weller, G.~Tabor, H.~Jasak, and C.~Fureby.
\newblock {A tensorial approach to computational continuum mechanics using
  object-oriented techniques}.
\newblock {\em Comput. Phys.}, 12(6):620--631, 1998.

\bibitem{williams1955}
M.~L. Williams, R.~F. Landel, and J.~D. Ferry.
\newblock {The Temperature Dependence of Relaxation Mechanisms in Amorphous
  Polymers and Other Glass-forming Liquids}.
\newblock {\em J. Am. Chem. Soc.}, 77(14):3701--3707, 1955.

\bibitem{xue1995}
S.-C. Xue, N.~Phan-Thien, and R.~Tanner.
\newblock {Numerical study of secondary flows of viscoelastic fluid in straight
  pipes by an implicit finite volume method}.
\newblock {\em J. Non-Newton. Fluid Mech.}, 59(2-3):191--213, 1995.

\bibitem{xue2004}
S.-C. Xue, R.~Tanner, and N.~Phan-Thien.
\newblock {Numerical modelling of transient viscoelastic flows}.
\newblock {\em J. Non-Newton. Fluid Mech.}, 123(1):33--58, 2004.

\bibitem{yesilata2000}
B.~Yesilata, A.~{\"{O}}ztekin, and S.~Neti.
\newblock {Non-isothermal viscoelastic flow through an axisymmetric sudden
  contraction}.
\newblock {\em J. Non-Newton. Fluid Mech.}, 89(1-2):133--164, 2000.

\end{thebibliography}

\end{document}